\documentclass[fleqn,usenatbib]{mnras}

\usepackage{newtxtext,newtxmath}

\usepackage[T1]{fontenc}

\DeclareRobustCommand{\VAN}[3]{#2}
\let\VANthebibliography\thebibliography
\def\thebibliography{\DeclareRobustCommand{\VAN}[3]{##3}\VANthebibliography}

\usepackage{graphicx}	
\usepackage{subcaption} 
\graphicspath{ {./images/} }

\usepackage{amsmath}	
\usepackage{amssymb}    
\usepackage{bm}         

\title[Vortex-flux-tube pinning with proton feedback]{Stability of interlinked neutron vortex and proton flux-tube arrays in a neutron star -- III. Proton feedback}

\author[K. H. Thong, A. Melatos and L. V. Drummond]{
K. H. Thong$^{1,2}$
\thanks{E-mail: kokhongt@student.unimelb.edu.au}
,
A. Melatos$^{1,2}$
\thanks{E-mail: amelatos@unimelb.edu.au}
and
L. V. Drummond$^{3}$
\thanks{E-mail:
lisadrum@mit.edu}
\\
$^{1}$School of Physics, University of Melbourne, Parkville, Victoria, 3010 Australia\\
$^{2}$OzGrav, Australian Research Council Centre of Excellence for Gravitational Wave Discovery, University of Melbourne, Parkville, Victoria, 3010 Australia\\
$^{3}$Department of Physics and MIT Kavli Institute, MIT, Cambridge, MA 02139 USA
}

\date{Accepted XXX. Received YYY; in original form ZZZ}

\pubyear{2022}

\begin{document}
\label{firstpage}
\pagerange{\pageref{firstpage}--\pageref{lastpage}}
\maketitle

\begin{abstract}
The coupled, time-dependent Gross-Pitaevskii and Ginzburg-Landau equations are solved simultaneously in three dimensions to investigate the equilibrium state and far-from-equilibrium, spin-down dynamics of an interpenetrating neutron superfluid and proton type-II superconductor, as an idealized description of the outer core of a neutron star. The simulations generalize previous calculations without the time-dependent Ginzburg-Landau equation, where proton feedback is absent. If the angle $\theta$ between the rotation and magnetic axes does not equal zero, the equilibrium state consists of geometrically complicated neutron vortex and proton flux-tube tangles, as the topological defects pin to one another locally but align with different axes globally. During spin down, new types of motion are observed. For $\theta = 0$, entire vortices pair rectilinearly with flux tubes and move together while pinned. For $\theta \neq 0$, vortex segments pair with segments from one or more flux tubes, and the paired segments move together while pinned. The degree to which proton feedback impedes the deceleration of the crust is evaluated as a function of $\theta$ and the pinning strength, $\eta$. Key geometric properties of vortex-flux-tube tangles, such as filament length, mean curvature, and polarity are analyzed. It is found that proton feedback smooths the deceleration of the crust, reduces the rotational glitch sizes, and stabilizes the vortex tangle dynamics. The dimensionless control parameters in the simulations are mutually ordered to match what is expected in a real neutron star, but their central values and dynamic ranges differ from reality by many orders of magnitude due to computational limitations.  
\end{abstract}




\section{Introduction}

The outer core of a neutron star, with mass density $1.6 \la \rho / (10^{14} \text{ g cm}^{-3})\la 3.9$, is filled with three interpenetrating and interacting fluids: superfluid neutrons, superconducting protons and viscous electrons \citep{Baym_1969,Yakovlev_1999,Chamel_2008}. The superfluid neutrons corotate approximately but not exactly with the crust. They are threaded by $N_{\rm v} \sim 10^{16} (\Omega_{\rm c} /1 \, {\rm rad \, s^{-1}})$ vortices, each carrying a quantum of circulation $\pi \hbar/m_n = 1.98 \times 10^{-3} \text{ cm}^2 \text{ s}^{-1}$, so that the superfluid mimics rigid-body rotation on a macroscopic scale. As the crust of the neutron star spins down from electromagnetic braking, neutron vortices move outwards from the outer core, transferring angular momentum from the condensate to the crust. The superconducting protons are believed to be in a type-II state from a naive comparison of the proton coherence length $\xi_p$ and London penetration depth $\lambda_p > \xi_p/\sqrt{2}$ \citep{Baym_1969, Pines_1985}, although other states are possible in some regions, depending on the details of the anisotropic neutron and proton interactions \citep{Buckley_2004, Alford_2008,haber_critical_2017,Wood_2022}. The type-II superconducting protons corotate with the crust. They are threaded by $N_{\rm f} \sim 10^{30} (B/10^{12}\text{ G})$ flux tubes, each carrying a magnetic flux quantum $\pi \hbar/2e = 2.07 \times 10^{-7} \text{ G cm}^2$. The neutron and proton topological defects pin strongly to each other via density and current-current coupling with typical interaction energy $\sim$ MeV per junction \citep{Scrinivasan_1990}. Hence the neutron dynamics within the outer core are influenced by the proton flux-tube array, even though the proton density is $\lesssim 5\%$ of the neutron density \citep{chau_implications_1992}.

It is natural to ask to what extent the proton flux tubes interfere with the outward migration of neutron vortices, and vice versa, as the crust spins down in response to an external torque. (i) Do the vortices pin to the flux tubes, so that their migration is arrested, and the deceleration of the neutron superfluid stalls? (ii) Does the back-reaction on the protons modify the flux-tube array, e.g. by driving instabilities which disrupt its unstressed rectilinear form? In previous numerical studies, \citet{drummond_2017} addressed question (i) in the neutron star context by modelling the dynamics of the neutron superfluid with the Gross-Pitaevskii equation (GPE) in the presence of a static flux-tube array, with zero back-reaction on the protons. \citet{drummond_2017} found three key results: (a) a glassy approach to equilibrium through a quasistatic sequence of metastable states; (b) neutron vortex tangles characteristic of quantum turbulence \citep{barenghi_quantized_2001}; and (c) accelerated spin down of the superfluid, and hence decelerated spin down of the crust, because tangled neutron vortices slip past rectilinear proton flux tubes more easily than rectilinear neutron vortices \citep{drummond_2018}. Neutron vortices tangle, because vortex-vortex repulsion competes with vortex-flux-tube attraction in a complicated fashion, when the rotation and magnetic dipole axes are misaligned. 

The simulations in \citet{drummond_2017} and \citet{drummond_2018} do not address question (ii) above, as the protons in those simulations are static, not dynamic; they are unable to respond to the neutrons, e.g. the forces exerted by the pinned neutron vortices. In this paper, we extend the simulations in \citet{drummond_2017} and \citet{drummond_2018} by allowing the protons to respond self-consistently to the motion of the neutrons, and vice versa. That is, we solve the proton Ginzburg-Landau equation (GLE) simultaneously with the neutron GPE in the presence of density coupling. The simulations are performed in imaginary time first, to study equilibrium configurations as in \citet{drummond_2017}, and then in real time, to study the response to secular electromagnetic spin down of the crust, as in \citet{drummond_2018}. The work is motivated by structural questions like the following. Do the proton flux tubes tangle microscopically? If they do, do neutron vortices slip past tangled proton flux tubes more easily during spin down, or is the opposite true, because the proton flux tubes keep pace with the neutron vortices and stay pinned? Are the sizes and rate of glitches suppressed, because proton flux tubes move with neutron vortices, or are glitches more frequent, because tangled flux tubes destabilize the system? 

We emphasize at the outset that the numbers of vortices and flux tubes studied in this paper are significantly less than the expected numbers of vortices and flux tubes in neutron stars due to computational limitations. Likewise, other dimensionless control parameters in the simulations (such as the proton fraction, neutron-proton coupling, pinning site separation, and London penetration depth, to name a few) are mutually ordered to match what is expected in a real neutron star, but their central values and dynamic ranges differ from reality by many orders of magnitude. In addition, the GPE and GLE assume dilute matter near its transition temperature and represent highly idealized descriptions of an interpenetrating neutron superfluid and proton superconductor at neutron drip densities, motivated again by computational cost. Overall, therefore, the calculations in this paper should be viewed as illustrative on a qualitative level; they do not represent a quantitative model of a neutron star. Extra caution should be exercised when extrapolating by analogy from the microscopic dynamics of vortex and flux-tube tangles to macroscopic astrophysical phenomena such as rotational glitches.

The paper is structured as follows. Section 2 outlines the equations of motion and procedure we use to numerically simulate the neutron superfluid, the proton superconductor, a crust experiencing an external torque, and their interactions with each other. Section 3 investigates the geometry of the different types of vortex-flux-tube junctions in equilibrium, before the spin-down torque is applied. Section 4 quantifies the effects of the dynamical proton fluid on the spin down of the crust. This includes comparing static and dynamic protons, different coupling strengths between the neutron and proton condensates, and different angles $\theta$ between the rotation and magnetic axes. Section 5 presents a geometric analysis of the vortex and flux-tube tangles. Properties such as filament length, mean curvature, and polarity are analyzed to deduce their impact on astrophysical observables, such as the spin-down rate of the crust.

\section{Equations of Motion}
In this section, we write down and justify physically the equations of motion for the various stellar components: neutrons (Section \ref{2.1}), protons (Section \ref{2.2}), neutron-proton interactions (Section \ref{2.3}), and the rigid crust (Section \ref{2.4}). The equations are solved numerically using the method described in Section \ref{2.5}.

The interaction between the neutron superfluid and proton superconductor in neutron stars has been studied in the microscopic regime via the GPE and GLE previously \citep{Alford_2008,haber_critical_2017,drummond_2017,drummond_2018,sauls_superfluidity_2019,Wood_2022}. The GPE and GLE resolve individual vortices and flux tubes and hence are suitable for studying vortex flux-tube tangles, the focus of this paper. In the macroscopic regime, the coupled superfluid and superconductor have been studied extensively via magnetohydrodynamics, i.e.\ a continuum description which averages over vortices and flux tubes \citep{mendell_superfluid_1991,mendell_superfluid_1991-1,mendell_magnetohydrodynamics_1998,glampedakis_magnetohydrodynamics_2011,dommes_vortex_2017,gusakov_force_2019,andersson_relativistic_2021}. \citet{gusakov_relativistic_2016} and \citet{sourie_vortex_2021} generalized the magnetohydrodynamic description to embrace general relativity.

\subsection{Neutrons}
\label{2.1}
We model the neutron superfluid in the outer core using the GPE. This is an idealization, of course. The GPE applies to dilute, weakly interacting particles near the transition temperature, whereas the neutrons are dense and strongly interacting and are colder than the transition temperature in the outer core of neutron stars $\gtrsim 1 \, {\rm week}$ old \citep{pethick_cooling_1992, Yakovlev_1999}. Nevertheless, the GPE naturally leads to vortex formation, which is the central concern of this paper. It captures broadly the dynamics of a neutron condensate, especially when considering the lack of specific experimental information about the physical conditions (e.g. thermodynamic phases) inside a neutron star. It has been employed previously to reproduce successfully the sporadic dynamics and size and waiting-time statistics of neutron star glitches caused by vortex avalanches \citep{warszawski_grosspitaevskii_2011, melatos_persistent_2015, haskell_models_2015, lonnborn_collective_2019}. For more discussion on the advantages and limitations of this model, the reader is referred to Section 5 of \citet{drummond_2017} and Section 7 of \citet{drummond_2018}. 

The GPE in a uniformly rotating frame with angular velocity $\mathbf{\Omega}_{\rm c}$ can be derived by minimizing the free energy functional of the neutron condensate \citep{pethick_boseeinstein_2008, eysden_van_anthony_superfluid_2011},
\begin{align}
\label{eq:2.1}
    F_n = &\int d^3x \ \biggl[ \frac{\hbar^2}{2m} |\nabla \psi|^2 -\frac{\mu_n}{m} |\psi|^2 - V_{\omega} |\psi|^2 + \frac{U_0}{2} |\psi|^4 \nonumber \\ 
    & - (\mathbf{\Omega}_{\rm c}\times\mathbf{x}) \cdot \mathbf{j}_n  \biggr],
\end{align}
where the neutron probability current is given by
\begin{equation}
    \mathbf{j}_n = \frac{i\hbar}{2}(\psi \nabla \psi^* - \psi^* \nabla \psi).
\end{equation}
In (\ref{eq:2.1}), $\psi$ is the complex neutron order parameter normalized by the number $N_n = \int{d^3x \ |\psi|^2}$ of neutron Cooper pairs, $m = 2m_n$ is the mass of the neutron Cooper pair, $\mu_n$ is the chemical potential of the neutron condensate, $V_{\omega}$ is the harmonic potential trapping the condensate (here, the gravitational potential of the star), and $U_0$ is the neutron self-interaction strength. We trap the neutrons using a harmonic potential with cylindrical symmetry, given by $V_\omega = \omega^2(x^2 + y^2)/2$, where $\omega$ is the trap frequency, which corresponds gravitationally to the idealized approximation of a constant-density star.

We follow previous papers \citep{melatos_persistent_2015,drummond_2017,drummond_2018} and solve the dimensionless stochastic GPE from \citet{gardiner_stochastic_2002},
\begin{align}
\label{eq:2.3}
    (i - \gamma) \frac{\partial \psi}{\partial t} =& \biggl(-\frac{1}{2} \nabla^2 + V_\omega + |\psi|^2 - \Omega_{\rm c} \hat{L}_z  + i\gamma \mu'_n \biggr) \psi + \mathcal{H}_{\rm int}[\psi, \phi],
\end{align}
where the rotation occurs about the $\hat{z}$-axis without loss of generality, $\hat{L}_z$ is the angular momentum operator, $\mu'_{n} \neq \mu_n$ is the chemical potential of the uncondensed neutrons, $\gamma > 0$ is a real damping rate, and $\mathcal{H}_{\rm int}[\psi, \phi]$ is the neutron-proton interaction Hamiltonian. Equation (\ref{eq:2.3}) is dimensionless with characteristic length-scale and time-scale given by the neutron coherence length $\xi_n = \hbar/(2m_n n_n U_0)^{1/2}$ and $\tau_n = \hbar/(n_n U_0)$ respectively, where $n_n$ is the neutron mean number density. The $-\gamma \partial \psi / \partial t$ term in the stochastic GPE models dissipation from realistic processes such as particle exchange between the superfluid and normal neutrons, mutual friction between the vortices and normal matter, and inelastic phonon emissions from the condensed neutrons \citep{mendell_superfluid_1991,mendell_superfluid_1991-1}. It acts as an approximate numerical device to stabilize vortex-array formation, which cannot be done with the undamped GPE derived from equation (\ref{eq:2.1}) \citep{tsubota_vortex_2002,gardiner_stochastic_2002}. Dissipation drives particle loss and hence is counterbalanced by the $i \gamma \mu'_{n} \psi$ term, which couples the neutrons to a thermal reservoir (e.g. a cloud of uncondensed neutrons) and damps sound waves.

\subsection{Protons}
\label{2.2}
We model the proton superconductor in the outer core using the GLE with the standard minimal electromagnetic coupling $\nabla \mapsto \nabla - iq \mathbf{A}/(\hbar c)$, where $q = 2e$ is the charge of a proton Cooper pair, and $\mathbf{A}$ is the magnetic vector potential. The GLE is another idealization. It applies to materials near the transition temperature (again, a pragmatic assumption to enable progress, which is not always true in a neutron star) and it is restricted to spatial variations that are not too rapid \citep{tinkham2004introduction}. The time-independent GLE in a uniformly rotating frame with angular velocity $\mathbf{\Omega_{\rm c}}$ can be derived by minimizing the free energy functional of the proton condensate \citep{tinkham2004introduction},
\begin{align}
\label{eq:2.4}
    F_{p} = &\int d^3x \ \biggl[ \alpha |\phi|^2 - V|\phi|^2 + \frac{\beta}{2} |\phi|^4 + \frac{\hbar^2}{2m} \biggl| \biggl( \nabla - \frac{iq}{\hbar c} \mathbf{A}\biggr) \phi \biggr|^2 \nonumber \\
    & + \frac{1}{8\pi}|\nabla \times \mathbf{A}|^2 - (\mathbf{\Omega}_{\rm c}\times\mathbf{x})\cdot \mathbf{j}_p \biggr],
\end{align}
where the proton current density is given by
\begin{equation}
    \mathbf{j}_p = \frac{i\hbar}{2} \biggl[\phi \biggl( \nabla + \frac{iq}{\hbar c} \mathbf{A}\biggr) \phi^* - \phi^* \biggl( \nabla - \frac{iq}{\hbar c} \mathbf{A}\biggr) \phi \biggr].
\end{equation}
In (\ref{eq:2.4}), below the transition temperature, $\alpha(T) < 0$ and $\beta(T) > 0$ are real constants whose values depend on the temperature $T$ \citep{tinkham2004introduction}. The total potential $V$ is the sum of the harmonic trap $V_\omega$ and a grid of cylindrical ``pinning'' sites for the protons with potential $V_{\rm f}$. The pinning sites represent an artificial numerical device to implement flux freezing between the protons (which are locked to the crust astrophysically) and the magnetic field $\mathbf{B} = \nabla \times \mathbf{A}$. Their form and purpose are discussed further in Section \ref{2.4}. The complex proton order parameter $\phi$ is normalized by the number of proton Cooper pairs $N_p = f N_n = \int{d^3x \ |\phi|^2}$, where $f$ is the proton fraction. In an astrophysical setting, one has $f \lesssim 0.05$ typically in the outer core \citep{chau_implications_1992}. In this paper, we set $f \sim 1$ artificially, to ensure that the neutron and proton condensates have similar diameters and overlap substantially within the simulation volume (see Appendix \ref{appendix:a}). The electric field induced by Faraday's law and associated with the displacement current is neglected in (\ref{eq:2.4}), because we assume that wave-like electromagnetic effects propagate on the light crossing time-scale, which is much shorter than the slow time-scale in (\ref{eq:2.4}). The fluid in the outer core achieves charge neutrality faster than the slow time-scale in (\ref{eq:2.4}) due to the highly electrically conductive relativistic electrons.

The dimensionless time-dependent GLE (TDGLE) solved here is an equation with diffusive \citep{schmid_time_1966,kopnin_introduction_2002, tinkham2004introduction} and wave-like \citep{ebisawa_wave_1971} properties which models the dynamics of a superconductor away from equilibrium. The TDGLE in the frame rotating with angular velocity $\mathbf{\Omega}_{\rm c}$ takes the form
\begin{align}
\label{eq:2.6}
    -\Gamma \frac{\partial \phi}{\partial t} =& \biggl[\biggl(-\frac{i}{\kappa_{\rm GL}}\nabla - \mathbf{A}\biggr)^2 + V - 1 + |\phi|^2 - \Omega_{\rm c} \hat{L}_z\biggr]\phi + \mathcal{H}_{\rm int}[\phi, \psi], 
\end{align}
in the electromagnetic gauge with zero scalar potential, where $\mathcal{H}_{\rm int}[\psi, \phi]$ is the neutron-proton interaction Hamiltonian, $\kappa_{\rm GL} = \lambda_p/\xi_p$ is the Ginzburg-Landau parameter, and $\xi_p$ and $\lambda_p$ are the characteristic lengths of $\phi$ and $\mathbf{A}$ respectively. Equation (\ref{eq:2.6}) is dimensionless with length-scale and time-scale given by $\lambda_p$ and $\tau_p$ respectively. In this paper, we assume $\lambda_p \sim \xi_n$ \citep{mendell_superfluid_1991} and $\tau_p \sim \tau_n$; that is, for illustrative purposes, we assume that the diffusive and wave-like time-scales of the neutrons and protons are similar in the absence of information to the contrary and choose $\Gamma = \gamma - i$.

The rotational evolution takes place over the spin-down time-scale $\gtrsim 10^3 \, {\rm yr}$ in a neutron star, as does the concomitant rotation-driven evolution of the magnetic field, e.g.\ due to vortex-flux-tube coupling. Magnetic field evolution due to Ohmic dissipation or Hall drift takes even longer, e.g.\ $\gtrsim 10^5 \, {\rm yr}$ \citep{pons_magnetic_2007,pons_magnetic_2019}. By comparison, physical effects on the light-crossing time-scale, such as the displacement current, are neglected. Astrophysically, therefore, it is a good approximation to hold $\mathbf{A}$ constant in this paper and evolve the proton fluid under a mean magnetic field $\overline{\mathbf{B}} = \nabla \times \overline{\mathbf{A}}$. This agrees qualitatively with \citet{Baym_1969}, where the protons transition into the superconducting state under a frozen-in magnetic field because of the high conductivity of the relativistic electrons in the outer core. 

For $B \sim 10^{12} \text{ G}$, the distance between flux tubes, $d_{\rm f}$, satisfies $\lambda_p \sim 10^{-12} \text{ cm} \ll d_{\rm f} \sim 10^{-10} \text{ cm}$ \citep{Baym_1969} in the Abrikosov ground state. However, because flux tubes pin to vortices, flux tubes may approach closer to each other than their triangular lattice separations. \citet{Scrinivasan_1990} estimated $d_{\rm f} \gtrsim 2 \times 10^{-11} \, {\rm cm}$, at which point magnetic interactions become important locally. Further, the neutron fluid entrains protons. Flux tubes each have a magnetic field $B_p \sim 10^{15} \, {\rm G}$. The entrained protons generate a magnetic field $B_n \sim 10^{14} \, {\rm G}$ and interact with other vortices and flux tubes magnetically. Nevertheless, as discussed in Appendix \ref{appendix:b}, we keep $\mathbf{A} = \overline{\mathbf{A}}$ to render the calculations in this paper tractable and leave the evolution of $\mathbf{A}$ via local magnetic interactions and entrainment to future work. 

\subsection{Multiple fluids}
\label{2.3}
The neutron and proton fluids interact via density and entrainment effects \citep{srinivasan_novel_1990, bhattacharya_evolution_1991}. To model their dynamics with their interactions included, we minimize the total free energy functional
\begin{equation}
\label{eq:2.7}
    F = F_n + F_p + F_{\rm int}.
\end{equation}
In (\ref{eq:2.7}), $F_{\rm int}$ is the dimensional interaction term \citep{alpar_rapid_1984, Alford_2008, drummond_2017} given by
\begin{equation}
    F_{\rm int} = \int{d^3x \ \biggl(\varsigma |\psi|^2 |\phi|^2 + \zeta \mathbf{j}_n \cdot \mathbf{j}_p \biggr)},
\end{equation}
where $\varsigma$ and $\zeta$ are dimensional density and current-current coupling constants respectively. \citet{drummond_2017} took the dimensionless density coupling constant to be $\eta = \varsigma/U_0 \approx -0.2$ for typical neutron star parameters. For simplicity, in this paper, we only study density interactions ($\varsigma \neq 0, \ \zeta = 0)$. As was seen in \citet{drummond_2018}, density interactions are sufficient to begin exploring the central questions addressed in this paper, such as the nature of spin-down dynamics in the presence of proton feedback. We leave the inclusion of current-current interactions to future work. For related work with current-current interactions, the reader is referred to \citet{drummond_2017}. Minimizing the dimensionless $F_{\rm int}$ with respect to $\psi^*$ and $\phi^*$, we get 
\begin{align}
    \mathcal{H}_{\rm int}[\psi, \phi] &= \eta |\phi|^2 \psi, \\ 
    \mathcal{H}_{\rm int}[\phi, \psi] &= \eta |\psi|^2 \phi 
\end{align}
respectively in equations (\ref{eq:2.3}) and (\ref{eq:2.6}). 

\subsection{Crust}
\label{2.4}
Equations (\ref{eq:2.3}) and (\ref{eq:2.6}) are expressed in the rotating frame of the crust. Following previous authors \citep{warszawski_grosspitaevskii_2011, melatos_persistent_2015, drummond_2018, howitt_simulating_2020}, we simulate the spin down of the crust under the action of an electromagnetic braking torque, $N_{\text{em}}$, which is constant to a good approximation over the time-scale of interest, because the waiting time between glitches is much shorter than the spin-down time-scale. We incorporate the back-reaction of the condensates with a dimensionless equation of the form
\begin{equation} 
\label{eq:2.11}
    L_{\rm tot}(t) = L_{\rm tot}(t=0) - N_{\rm em} t,
\end{equation}
where $L_{\rm tot}(t)$ is the total angular momentum of the system at time $t$ given by
\begin{equation}
\label{eq:2.12}
    L_{\rm tot}(t) = I_{\rm c}\Omega_{\rm c}(t) + \langle \psi(t) | \hat{L}_z | \psi(t) \rangle + \langle \phi(t) | \hat{L}_z | \phi(t) \rangle,
\end{equation}
and $I_{\rm c}$ is the dimensionless moment of inertia of the crust. 

We ``pin'' the proton flux tubes to an array of cylindrical potentials oriented parallel to $\overline{\mathbf{B}}$ and corotating with the crust. The cylindrical potentials are an artificial numerical device used to promote magnetic flux freezing, as noted in Section \ref{2.2}; they do not correspond to physical potential wells, such as nuclear lattice sites, because the simulations are done in the star's outer core, where there is no nuclear lattice by assumption. Their dimensionless form is given by
\begin{equation}
\label{eq:2.13}
    V_{\rm f} = -\sum_k{V_0 \exp \biggl[\frac{-(\mathbf{x}_\perp-\mathbf{x}_{\perp, k})^2}{r_0^2} \biggr]},
\end{equation}
where $V_0 > 0$ and $r_0 > 0$ are the characteristic depth and width of the pinning potential respectively, $k$ labels a pinning site, and the subscript $\perp$ denotes a displacement perpendicular to the axis of the cylinder, cf.  \citet{howitt_anti-glitches_2022}. The phenomenological expression (\ref{eq:2.13}) ensures that the protons and magnetic field are frozen together approximately and hence also corotate with the crust, which is locked electromagnetically to the charged fluid in a neutron star. We are obliged to introduce this artificial numerical device, because the magnetohydrodynamic property of flux freezing arises physically from the interplay between Faraday's law and the high electrical conductivity of the electrons in Amp\'{e}re's law \citep{baym_electrical_1969}, but we do not solve Faraday's law and Ampere's law in this paper to keep the problem tractable. In the inner crust, vortices also pin to physical inhomogeneities in the crustal lattice, which may be microscopic (e.g.\ nuclei or their interstices) \citep{link_mechanics_1991, link_superfluid_1993, seveso_mesoscopic_2016, avogadro_quantum_2007, link_vortex_2022} or mesoscopic (e.g.\ dislocations, cracks) \citep{chapman_vortex_1997,tinkham2004introduction,middleditch_predicting_2006,kerin_mountain_2022}, but physical pinning of this kind is excluded from this paper, because we focus on the outer core instead of the inner crust.

\subsection{Numerical method}
\label{2.5}
We solve equations (\ref{eq:2.3}) and (\ref{eq:2.6}) simultaneously using the fourth-order Runge-Kutta in the Interaction Picture (RK4IP) algorithm developed by \citet{caradoc-davies_vortex_2000}. Equations (\ref{eq:2.3}) and (\ref{eq:2.6}) are ordinary differential equations in the interaction picture, which can be solved by a variety of numerical methods. The fourth-order Runge-Kutta scheme is chosen because it has low memory consumption and reduces the number of computationally expensive diffusion exponentials. For more details on the RK4IP's optimization and implementation, the reader is referred to \citet{caradoc-davies_vortex_2000}. 

Following \citet{drummond_2017,drummond_2018}, we use imaginary time evolution ($t \rightarrow -it, \gamma \rightarrow 0$) to first find the equilibrium states of equation (\ref{eq:2.3}) and (\ref{eq:2.6}) while keeping $\Omega_{\rm c}$ fixed. We designate an equilibrium as being reached, when both the convergence metrics $\int{d^3x \ |\hat{\Delta} \psi|^2} \leq 10^{-9} N_n$ and $\int{d^3x \ |\hat{\Delta} \phi|^2} \leq 10^{-9} N_p$ are satisfied, where $\hat{\Delta}$ is the operator that represents the change between successive time steps. The equilibrium state of the neutrons is the lowest energy eigenstate with eigenvalue $E_0 = \mu'_n$ and time evolution $ \propto e^{-i\mu'_n t}$. Hence, we can find $\mu'_n$ from the equilibrium $\psi$ obtained from imaginary time evolution.

The equilibrium states and $\Omega_{\rm c}$ are evolved in real time in two steps. Step (i) involves turning on $V_{\rm f}$ through $V$ in (\ref{eq:2.6}) and allowing $|\hat{\Delta} \psi|^2$, $|\hat{\Delta} \phi|^2$, and $\hat{\Delta} \Omega_{\rm c}$ to converge. This allows neutrons and protons to find new equilibria with $V_{\rm f} \neq 0$ and when $\Omega_{\rm c}$ is no longer fixed, so neutrons and protons are allowed to transfer angular momentum to the crust. Step (ii) involves turning on $N_{\rm em}$ in (\ref{eq:2.11}) to spin down the crust to study the far-from-equilibrium dynamics. We allow $\Omega_{\rm c}$ to vary and $\hat{\Delta}\Omega_{\rm c}$ to converge first before step (ii), because the final state found from imaginary time evolution is not the true equilibrium state but a metastable low-energy state of the superfluid, which can relax further and transfer residual amounts of angular momentum to or from the crust. During steps (i) and (ii), the radii of the condensates change (arranged to be approximately equal for $f \sim 1$, as explained in Appendix \ref{appendix:a}, and denoted by $R$), because the centrifugal force decreases, as $\Omega_{\rm c}$ decreases. The angular momentum of the neutrons depends on $R$, e.g., $\langle \psi | \hat{L}_z | \psi \rangle \propto \sum_{i} (R^2 - r_i^2)$ is the dominant term for an axisymmetric superfluid \citep{shaham_superfluidity_1980}, where the sum is over all vortices, and $r_i$ is the cylindrical distance of the $i$-th vortex from the axis of the condensate. Hence changes in $R$ lead to changes in $\langle \psi | \hat{L}_z | \psi \rangle$ and $d\Omega_{\rm c}/dt$ via (\ref{eq:2.11}). In this paper, we do not seek to study how $R$ changes, as the size of the condensate is set by other factors (chiefly stratification of thermodynamic phases) in a neutron star, so we keep $R$ fixed. To do this, we note that $V_{\omega}$ has the same form as the centrifugal potential. Hence we modify $V_{\omega}$ continuously to counterbalance changes in the centrifugal potential $\propto \Omega_{\rm c}^2$, cf. \citet{drummond_2018} where $V_{\omega}$ is fixed.

\section{Vortex-flux-tube intersections}
\label{sec:3}
\begin{figure}
    \begin{subfigure}[b]{0.24\textwidth}
    	\includegraphics[width=\textwidth]{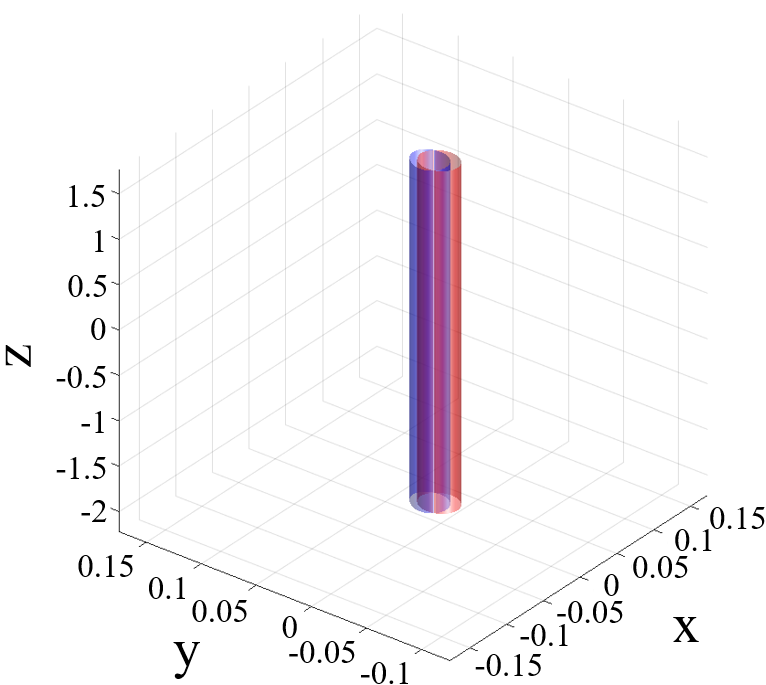}
    	\caption{}
    	\label{fig:1a}
    \end{subfigure}
    \hfill
    \begin{subfigure}[b]{0.24\textwidth}
    	\includegraphics[width=\textwidth]{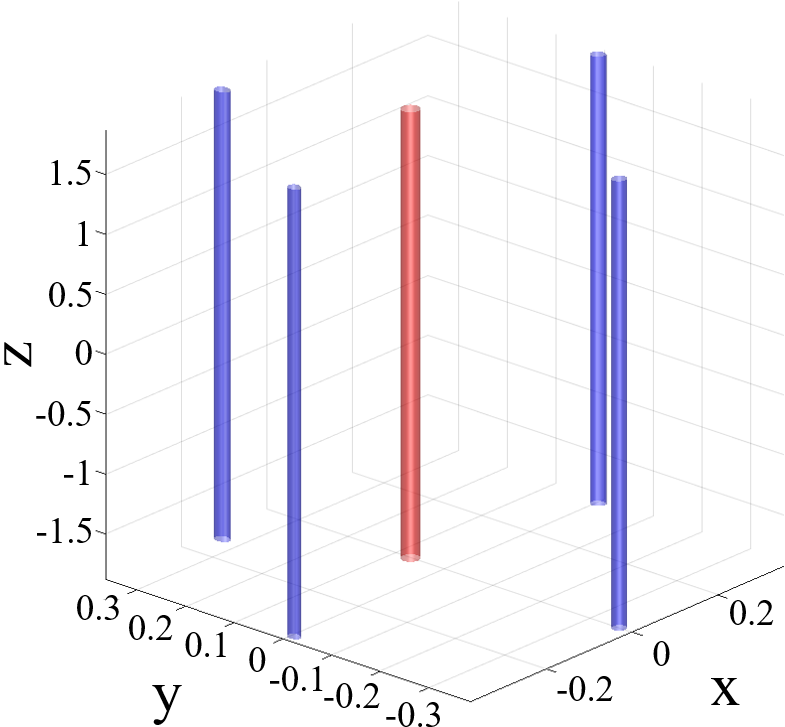}
    	\caption{}
    	\label{fig:1b}
    \end{subfigure}
    \hfill
    \begin{subfigure}[b]{0.24\textwidth}
    	\includegraphics[width=\textwidth]{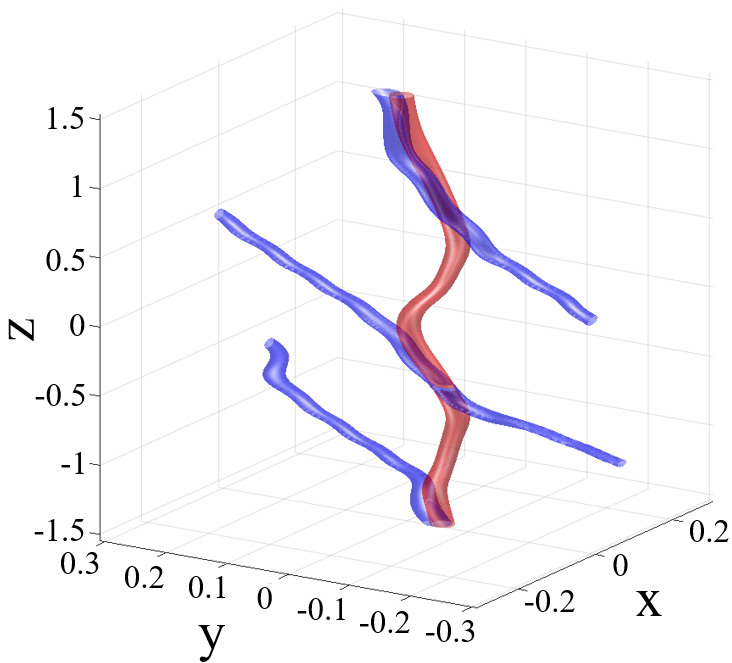}
    	\caption{}
        \label{fig:1c}
    \end{subfigure}
    \hfill
    \begin{subfigure}[b]{0.24\textwidth}
    	\includegraphics[width=\textwidth]{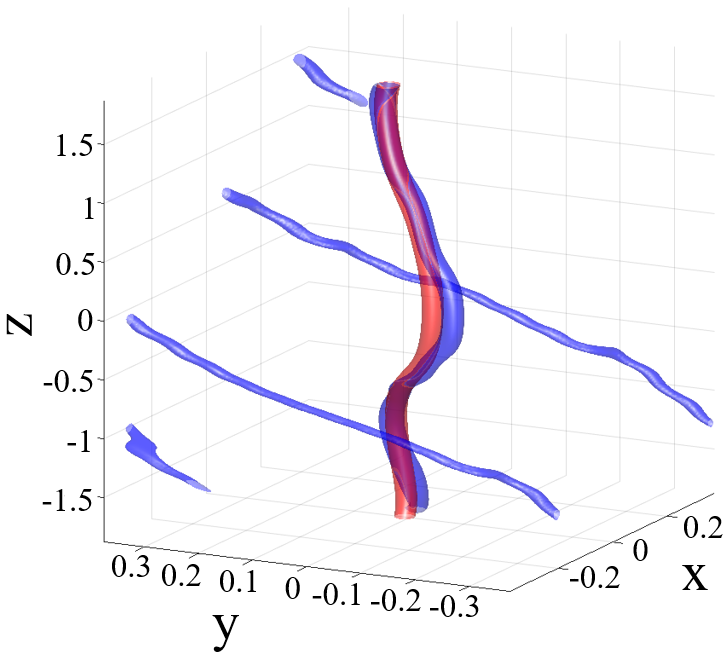}
    	\caption{}
        \label{fig:1d}
    \end{subfigure}
    \hfill
    \begin{subfigure}[b]{0.24\textwidth}
    	\includegraphics[width=\textwidth]{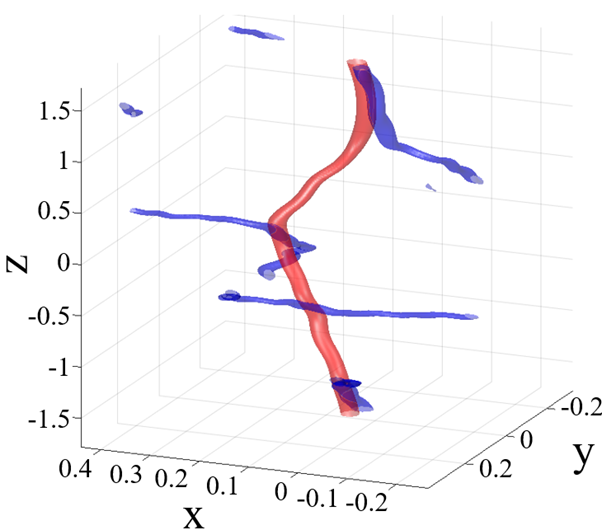}
    	\caption{}
        \label{fig:1e}
    \end{subfigure}
    \hfill
    \begin{subfigure}[b]{0.24\textwidth}
    	\includegraphics[width=\textwidth]{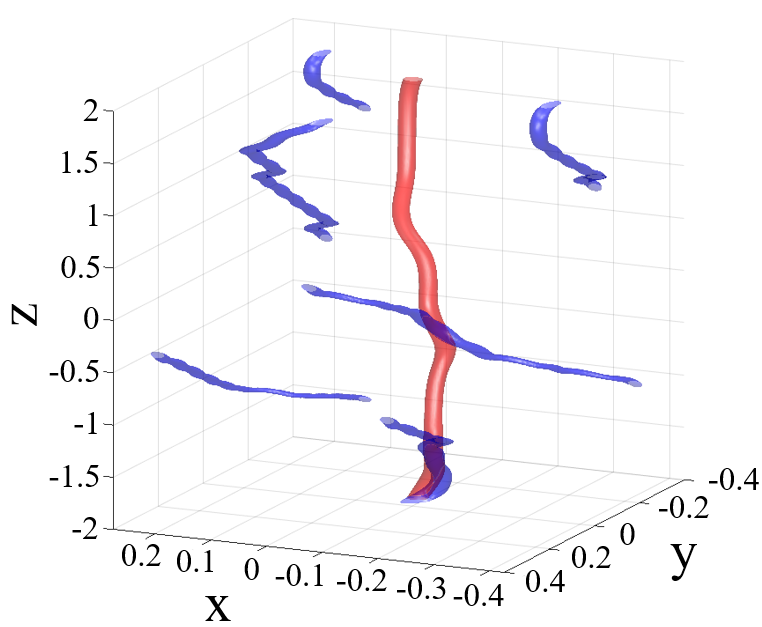}
    	\caption{}
        \label{fig:1f}
    \end{subfigure}
    \caption{Close-up snapshots in imaginary time of vortices (red shading) and flux tubes (blue shading), showing their overall deformation and the local geometry of the pinning junctions where they intersect. The shading signifies where  $|\psi|^2$ or $|\phi|^2$ drops to $0.05$ times its maximum. The first, second and third rows are for $\theta = 0, \, \pi/6, \, \pi/4$ respectively. The first and second columns are snapshots of two different and arbitrary vortices. Length units are $\xi_n$.  Parameters: $\Omega_{\rm c}(t=0)=4, \, |\overline{\mathbf{B}}|=8, \, \kappa_{\rm GL}=2, \, \omega=120, \, V_0 = -75, \, r_0^2 = 0.005, \, \eta=-0.5, \, \gamma=0, \, N_{\rm em}=0, \, \tilde{N}_n=N_n/(n_n \xi_n^3)=1.6\times10^4 \text{\ and} \ \tilde{N}_p=1.25\tilde{N}_n.$}
    \label{fig:1}
\end{figure}

We begin by investigating the different types and orientations of vortex-flux-tube intersections during imaginary time evolution. Vortices and flux tubes are regions of low density in their condensates. Density coupling makes them attract and intersect. The number, location and geometry of their intersections provide important insights into how the filaments behave in far-from-equilibrium scenarios, which we study in Section \ref{sec:4}. 

\subsection{Rotation and magnetic axes aligned}
\label{sec:3.1}
In imaginary time evolution, vortices and flux tubes form at the edge of the condensates and move inwards until they settle into the lowest energy configuration. The path to equilibrium depends on $\theta$. Vortices and flux tubes attract each other locally while aligning globally with the rotation and magnetic axes respectively \footnote{The flux tubes in this paper do not align perfectly with $\overline{\mathbf{B}}$ globally. Instead, they align with an axis that is inclined between $\overline{\mathbf{B}}$ and $\mathbf{\Omega}_{\rm c}$. This is caused by competition between the minimization of the fourth and last terms in equation (\ref{eq:2.4}), where the fourth term favours global alignment with $\overline{\mathbf{B}}$, while the last term favours proton angular momentum aligning with $\mathbf{\Omega}_{\rm c}$. This issue does not persist if we allow $\mathbf{A}$ to vary consistently by solving Maxwell's equations (see Appendix \ref{appendix:b}), a topic for future work. \label{footnote:2}}. For $\theta = 0$, the cores of vortices and flux tubes overlap rectilinearly, as in Figure \ref{fig:1a}. A vortex-flux-tube pair forms on the boundary. As the pair propagates inwards, the flux-tube is held back by the repulsion of surrounding flux tubes and pulls on the vortex. If it is energetically favourable for the pinned vortex to continue moving inwards, it unpins from its partner flux-tube simultaneously along its entire length and pins to another flux-tube closer to the centre of the condensate, hopping between flux tubes \citep{haskell_pinned_2016}. Figure \ref{fig:1b} demonstrates a snapshot of an unpinned vortex in the middle of hopping from one flux-tube to another flux-tube. We note that the motion of the rectilinear vortices in our study differs from the motion in other studies with static pinning sites \citep{warszawski_grosspitaevskii_2011, melatos_persistent_2015, drummond_2017, howitt_simulating_2020}. Vortices hop between static pinning sites, whereas in this study, a vortex-flux-tube pair may move together. In other words, in this paper, the pinning sites themselves can move, when an intersecting vortex and flux-tube move as a pinned pair.

In the case of strong pinning, we define $\nu$ to be the characteristic cylindrical distance travelled by a pinned vortex-flux-tube pair divided by the characteristic cylindrical distance travelled when a vortex hops between flux tubes. The ratio $\nu$ has important observational consequences, if vortex avalanches are the mechanisms behind neutron star glitches. For $\nu \ll 1$, the Magnus force exceeds the pinning force, and the angular momentum of the neutrons is transferred to the crust in discrete step-like releases, where the unpinning of vortices leads to proximity knock-on and avalanches \citep{warszawski_grosspitaevskii_2011, haskell_pinned_2016}. For $\nu \gg 1$, the angular momentum is continuously transferred from the neutron condensate to the crust as vortices gradually creep outwards. The Magnus force pushes the vortex-flux-tube pairs cylindrically outwards and causes them to move together without splitting them apart. This leads to smooth spin down with no avalanches triggered by proximity knock-on. The ratio $\nu$ depends in part on how quickly the neutrons and protons respond to each other, with $\tau_n = \tau_p$ in this paper. If, in contrast, one has $\tau_n \ll \tau_p$, then flux tubes are approximately static and one reverts to the results in previous works \citep{warszawski_grosspitaevskii_2011, melatos_persistent_2015, drummond_2017, howitt_simulating_2020}. If one has $\tau_n \gg \tau_p$, then outwards moving vortices are always accompanied by pinned flux tubes, and hence vortex avalanches induced by vortex unpinning become rare or impossible.

The picture above relates to type-II superconductors. \citet{haber_critical_2017} and \citet{Wood_2022} have predicted the existence of type-1.5 proton superconductivity in some regions of the outer core, where regions of Meissner exclusion (type I) and flux tubes alternate. Vortex-flux-tube pairs traverse these alternating regions to reach the crust during spin down. In type-1.5 superconductivity, flux tubes form bundles due to their long-range attraction, so the Magnus force acting on a vortex-flux-tube pair must exceed the attractive force for the pair to move from a type II to a type I region. The attractive threshold may be an additional trigger for proximity knock-on and vortex avalanches and requires further investigation. 

\subsection{Rotation and magnetic axes misaligned}
\label{sec:3.2}

\begin{figure}
	\includegraphics[width=\columnwidth]{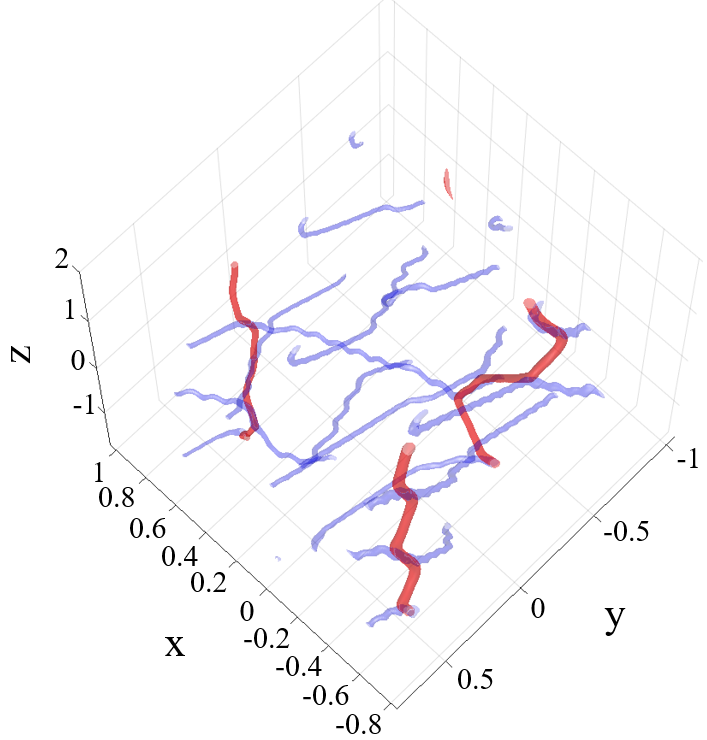}
	\caption{A zoomed out version of Figure \ref{fig:1e}, showing a larger simulation volume and viewed from a vantage point which displays clearly the complicated geometry of the vortex-flux-tube intersections locally and globally.}
	\label{fig:2}
\end{figure}

\begin{figure}
	\includegraphics[width=\columnwidth]{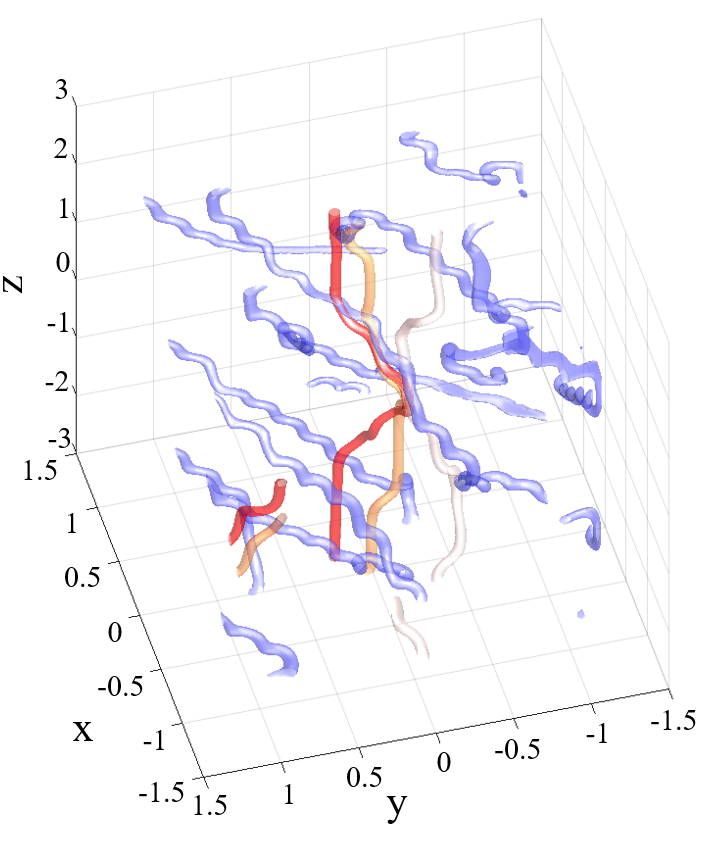}
	\caption{Close-up snapshots of a vortex slithering between flux tubes in imaginary time, i.e.\ during the approach to equilibrium. One vortex segment unpins then re-pins elsewhere on the same flux-tube, another vortex segment unpins then re-pins on a different flux-tube, and a third vortex segment never unpins during the imaginary time interval $1 \leq it \leq 3$. The blue shading signifies the flux tubes at $it = 2$. The milky white, salmon pink and red shading signifies the vortices at $it = 1, 2, 3$ respectively. Length units and paramters are the same as in Figure \ref{fig:1} except for: $\Omega_{\rm c}(t=0)=5, \, \theta = \pi/6, \, \omega=140, \, \tilde{N}_n=N_n/(n_n \xi_n^3)=4.8\times10^4 \text{\ and} \ \tilde{N}_p=1.25\tilde{N}_n.$}
	\label{fig:3}
\end{figure}

As $\theta$ increases from zero to $\pi/2$, vortices become less likely to only pin to a single flux-tube. Furthermore, vortices that pin to a single flux-tube are no longer rectilinear for $\theta \neq 0$. Figure \ref{fig:1d} illustrates a bent vortex-flux-tube pair for $\theta = \pi/6$. Most vortices pin along segments; that is, extended segments (not just single points) of the vortex coincide with extended segments of multiple nearby flux tubes, but the segments are disjoint, and the intervals between them are not pinned to anything. This partial pinning of vortices to flux tubes is clear in Figure \ref{fig:1c} ($\theta = \pi/6$), where three disjoint segments of a vortex (red filament) pin to three separate flux tubes (blue filaments) while aligning globally with the rotation axis. The same configuration is seen with the vortex in Figure \ref{fig:1e} ($\theta = \pi/4$) but the total length of the intersecting segments is smaller than in Figure \ref{fig:1c} because the rotation and magnetic axes are less aligned. Similarly, flux tubes pin partially to multiple nearby vortices by bending locally towards them while aligning globally with the magnetic axis. The competing tendencies locally and globally result in geometrically complicated vortex-flux-tube tangles, such as in Figure \ref{fig:2} ($\theta = \pi/4$; magnified image), where vortices and flux tubes intersect at multiple places. In Figure \ref{fig:2}, we observe that most flux tubes bend slightly towards nearby vortices, but the flux tube between the vortices at $(x, y) = (0.6, 0.4)$ and $(-0.2, -0.4)$ stretches considerably along the $x$ axis and makes a large angle $\approx \pi/6$ with $\overline{\mathbf{B}} = 8[\sin(\pi/4)\mathbf{\hat{y}} + \cos(\pi/4)\mathbf{\hat{z}}]$. We caution against inferring that flux-tube distortion changes the direction of $\overline{\mathbf{B}}$; we do not evolve $\overline{\mathbf{B}}$ in this paper.

For $\theta \neq 0$, the vortices ``slither'' rather than ``hop''. That is, one vortex segment unpins from a segment of a flux-tube then re-pins to another segment of the same or a different flux-tube, whilst other pinned segments of the same vortex remain pinned. Slithering differs from the hopping observed for $\theta = 0$, because slithering involves the partial unpinning of segments of vortices, while hopping involves the unpinning of vortices along their entire lengths. Slithering also differs from the ``zig-zagging'' motion from one static flux-tube to the next identified by \citet{drummond_2018}, because in this paper vortices and flux tubes can move together. The tendency to move together rises, as $|\eta|$ increases. Figure \ref{fig:1f} captures a snapshot of a vortex in the middle of slithering. The top end of the vortex unpins while the bottom end remains pinned. In Figure \ref{fig:1f}, the cores of the flux tubes at $z > 0$ do not overlap with the vortex core, but the flux tubes bend towards and attract the vortex, helping the vortex latch onto the next flux-tube as it slithers. After a vortex unpins from a segment of a flux-tube, it may re-pin to a different segment of the same flux-tube. We demonstrate this in Figure \ref{fig:3} with a close-up of a vortex at three different snapshots in imaginary time. The vortex segment at $z = 3$ unpins before re-pinning to a different segment of the same flux-tube, the segment at $z=-2$ unpins before re-pinning to a different flux-tube, and the segment at $z = 0$ remains pinned in place to another flux-tube segment. The flux tubes do not move appreciably during the event; only the vortices move between the three snapshots.

Does imaginary-time vortex slithering, as observed in Figure \ref{fig:3}, work the same way in real time, when the crust spins down? And does it enable the neutrons to transfer angular momentum more smoothly to the crust, and hence reduce the sizes and rate of glitches? In Section \ref{sec:4}, we study angular momentum transfer in real time with $N_{\rm em} \neq 0$ under a range of scenarios. The associated evolution of the vortex-flux-tube tangle (including its length, curvature, and polarity) is quantified in Section \ref{sec:5}.

\section{Spin down of the crust}
\label{sec:4}
In this section, we analyze how $\Omega_{\rm c}(t)$ and $\dot{\Omega}_{\rm c}(t)$ respond in real time, when the crust experiences a constant electromagnetic braking torque, chosen arbitrarily to be $N_{\rm em} = -0.05 I_{\rm c}$. We start from end states of imaginary time evolution with the same representative parameters as in Figure \ref{fig:3}, except with a range of $\eta$'s and $\theta$'s. We study how the crust spins down (i) with and without proton feedback (Section \ref{sec:4.1}), (ii) as a function of the neutron-proton pinning strength (Section \ref{sec:4.2}), and (iii) as a function of the inclination angle $\theta$ (Section \ref{sec:4.3}). We also calibrate against a non-astrophysical benchmark case, where the crust, neutrons, and protons corotate uniformly as a rigid body. 

In each numerical experiment, we identify and exclude from analysis an initial transient interval $0\leq t \leq t_{\rm ini} = 50$, with typically $\Omega_{\rm c}(t_{\rm ini}) \approx 0.75\Omega_{\rm c}(t=0)
$. We plot $\Omega_{\rm c}(t) - \Omega_{\rm c}(t_{\rm ini})$ for $t_{\rm ini} < t < t_{\rm end}$ where $t_{\rm end} = 90$ denotes when the experiment ends. We also plot the dimensionless spin-down rate $w_{\rm c}(t) = \dot{\Omega}_{\rm c}(t)/(-N_{\rm em}/I_{\rm tot})$ versus time and its probability density function (sampled in time steps $\Delta t=5\times 10^{-4}$ for $t > t_{\rm ini}$ during a single simulation run) as a histogram, while cautioning that the fluctuations are not necessarily stationary statistically. We  define the cumulative mean of the dimensionless spin-down rate as 
\begin{equation}
    \overline{w}_{\rm c}(t) = \frac{1}{t-t_{\rm ini}}\int_{t_{\rm ini}}^{t} \, dt' \, w_{\rm c}(t') ,
\end{equation}
and calculate $\overline{w}_{\rm c}(t)$ at different times during the simulation runs to study how the feedback onto the crust varies with time. 

\subsection{Angular velocity versus time with and without proton feedback}
\label{sec:4.1}
To quantify the effects of the proton feedback, we compare four scenarios: (1) neutrons and protons are attracted to each other ($\eta = \eta_{np} = \eta_{pn} \neq 0$), (2) neutrons are not attracted to protons ($\eta_{np} = 0$) but protons are attracted to neutrons ($\eta_{pn} \neq 0$), (3) neutrons are attracted to protons ($\eta_{np} \neq 0$) but protons are not attracted to neutrons ($\eta_{pn} = 0$), and (4) the crust, neutrons, and protons corotate as one rigid body. Only scenario (1) is realistic; the others are studied to develop physical intuition about causes and effects in the vortex and flux-tube dynamics. In scenario (2), the flux tubes tangle and pin to vortices but not vice versa, and hence the vortices are rectilinear. In scenario (3), the vortices tangle and pin to flux tubes but not vice versa. Scenario (3) resembles the work done by \citet{drummond_2018}, except that here the flux tubes adjust to changes in $\Omega_{\rm c}$ as they are not static.

\begin{figure}
    \begin{subfigure}[b]{\columnwidth}
    	\includegraphics[width=\columnwidth]{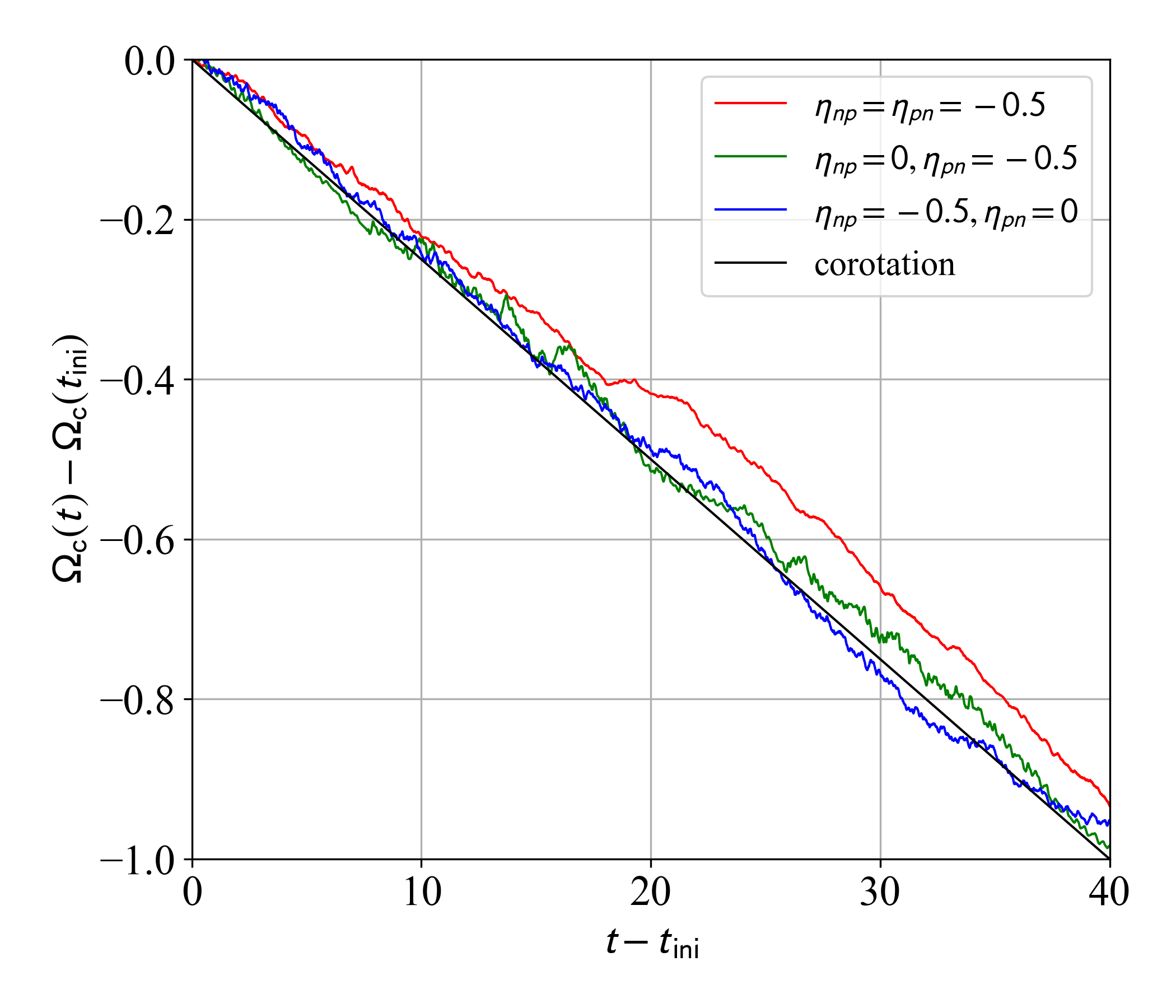}
    	\caption{}
    	\label{fig:4a}
    \end{subfigure}
    \hfill
    \begin{subfigure}[b]{\columnwidth}
    	\includegraphics[width=\columnwidth]{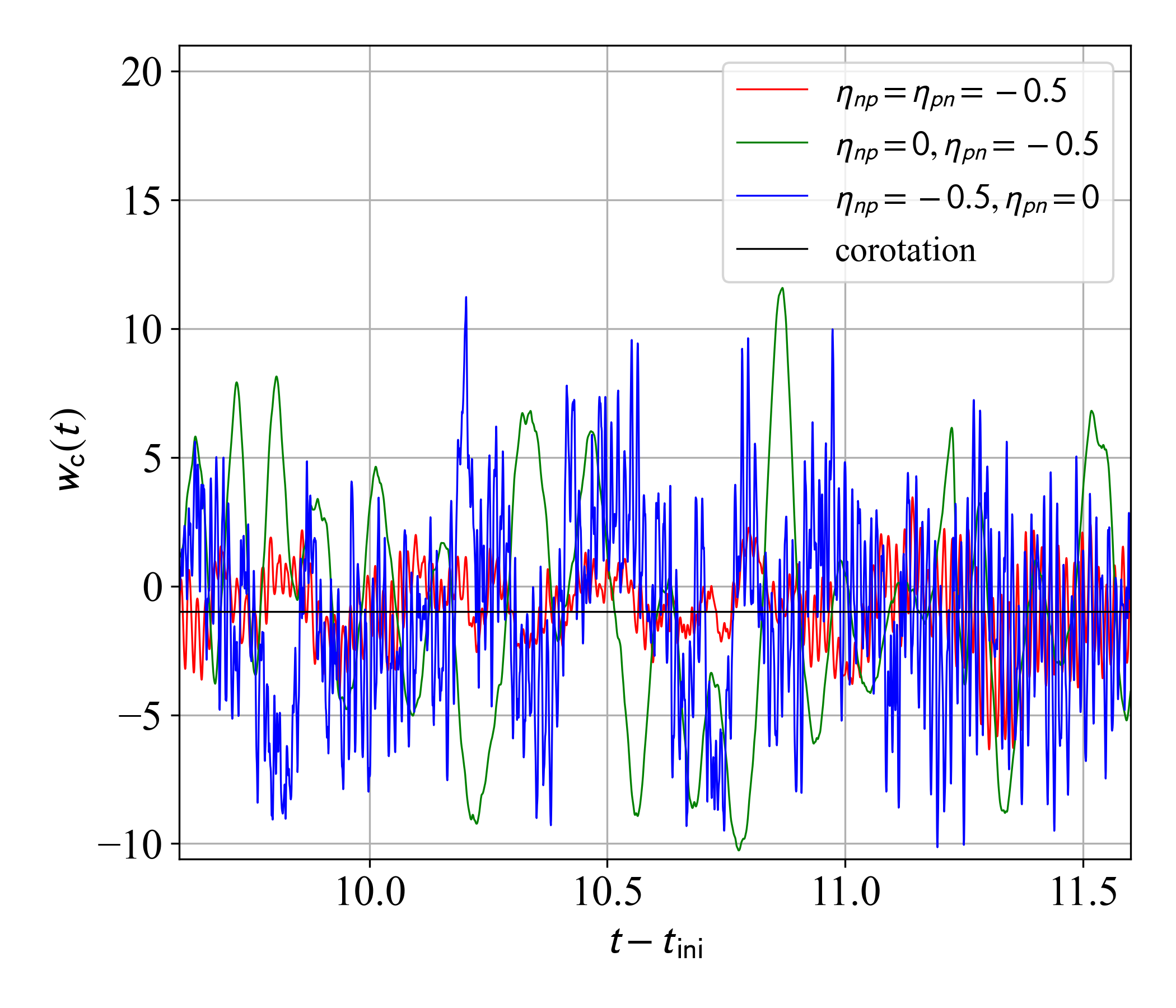}
    	\caption{}
    	\label{fig:4b}
    \end{subfigure}
    \hfill
    \begin{subfigure}[b]{\columnwidth}
    	\includegraphics[width=\columnwidth]{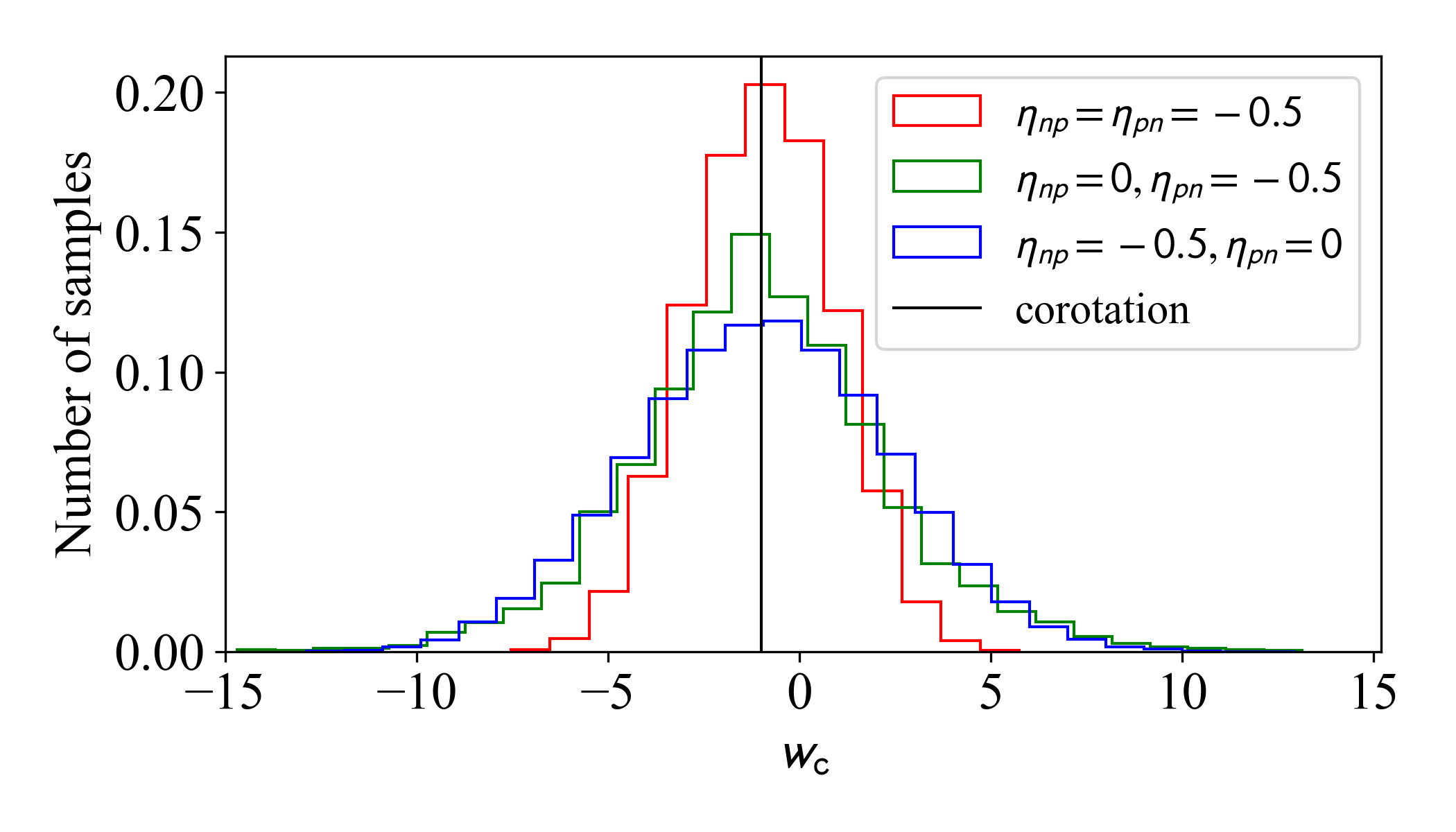}
    	\caption{}
    	\label{fig:4c}
    \end{subfigure}
    \caption{Rotation of the crust for the four combinations of $\eta_{np}$ and $\eta_{pn}$ specified in the legend. The black colour is for a corotating crust and core. (a) $\Omega_{\rm c}(t)  - \Omega_{\rm c}(t_{\rm ini})$ versus time. (b) $w_{\rm c}(t)$ versus time. (c) Probability density function of $w_{\rm c}$, sampled in steps of $\Delta t = 5\times 10^{-4}$ in the interval $t_{\rm ini} \leq t \leq t_{\rm end}$. Parameters: $t_{\rm ini} = 50, t_{\rm end} = 90, \Omega_{\rm c}(t=0)= 5, |\overline{\mathbf{B}}| = 8, \kappa_{\rm GL} = 2, \theta = \pi/6, \omega=140, V_{\rm f} = -75, \gamma=0.005, \tilde{N}_n=N_n/(n_n \xi_n^3)=4.8\times10^4 \text{\ and} \ \tilde{N}_p=1.25\tilde{N}_n.$}
    \label{fig:4}
\end{figure}

Figure \ref{fig:4} displays $\Omega_{\rm c}(t)  - \Omega_{\rm c}(t_{\rm ini})$, $w_{\rm c}(t)$, and the probability density function of $w_{\rm c}(t)$ for $t_{\rm ini} < t < t_{\rm end}$. The spin down of the crust in scenario (3) is similar to the spin down in scenario (4), as rectilinear vortices gradually creep outwards unimpeded in response to the spin down of the crust. We find that the spin down of the crust is retarded in scenario (1) compared to scenarios (2)--(4). Halfway through the post-transient interval, at $t_{\rm half} = (t_{\rm ini} + t_{\rm end})/2$, we obtain $\overline{w}_{\rm c}(t_{\rm half}) = -0.84, \, -1.0, \, -0.98, \, -1.0$ for scenarios (1)--(4), respectively (see Figure \ref{fig:4a}). The percentage difference between $\overline{w}_{\rm c}(t_{\rm half})$ in scenarios (1) and (3) is $[\overline{w}_{\rm c}^{(1)}(t_{\rm half}) - \overline{w}_{\rm c}^{(3)}(t_{\rm half})]/\overline{w}_{\rm c}^{(3)}(t_{\rm half}) = -15 \%$. This is consistent with proton feedback allowing neutron vortices to move more freely by slithering in scenario (1) (see Section \ref{sec:3.2}) as opposed to zig-zagging between the quasi-static pinning sites in scenario (3). 

As vortices annihilate at the crust, and $N_{\rm v}$ tends to zero (a limit which is not relevant astrophysically), the spin-down rates of the crust seems to plateau for all scenarios. By the end of the simulations, we obtain $\overline{w}_{\rm c}(t_{\rm end}) = -0.93,\, -0.98,\, -0.95,\, -1.0$ for scenarios (1)--(4), respectively (see Figure \ref{fig:4a}). The percentage difference between $\overline{w}_{\rm c}(t_{\rm end})$ in scenario (1) and (3) is small, $[\overline{w}_{\rm c}^{(1)}(t_{\rm end}) - \overline{w}_{\rm c}^{(3)}(t_{\rm end})]/\overline{w}_{\rm c}^{(3)}(t_{\rm end}) = -1.9\%$, as is the percentage difference between all other scenarios. Longer and larger simulations are needed to test whether the plateau and convergence are artifacts stemming from the non-astrophysical limit $N_{\rm v} \rightarrow 0$. However, such simulations are expensive computationally and lie outside the computational budget of this paper.

The transfer of angular momentum from the superfluid and superconductor to the crust varies appreciably between scenarios. In scenario (2), vortices closer to the centre of the condensate move outwards cylindrically to decelerate the superfluid but are pushed back by vortices further away from the centre. As a result, vortices oscillate radially in a slowly expanding Abrikosov lattice, causing $\langle \psi | \hat{L}_z | \psi \rangle$ and hence $\Omega_{\rm c}$ to oscillate via (\ref{eq:2.11}). These oscillations are seen in the blue curve in Figure \ref{fig:4b}. In scenarios (1) and (3), vortex segments also oscillate when repelled by nearby vortices or attracted by flux tubes where they pin. Both oscillation modes are driven by the Magnus force acting on the vortices. As the diameter of the flux-tube core is a lot smaller than the typical separation distance between vortices, vortex oscillation from vortex-vortex repulsion occurs with a lower frequency than the oscillation about flux tubes. This is reflected in Figure \ref{fig:4b}, where the oscillations of $w_{\rm c}(t)$ in scenario (2) have a lower frequency than the oscillations in scenarios (1) and (3). 

A key message from Figure \ref{fig:4b} is that proton feedback reduces the oscillation amplitude of both aforementioned modes. This implies that the rotation of the crust is smoother with proton feedback. Figure \ref{fig:4c} supports this argument. The histogram of $w_{\rm c}$ in scenario (1) is narrower than in scenarios (2) or (3). In the interval $t_{\rm ini} \leq t \leq t_{\rm end}$, the standard deviation $\sigma_{w_{\rm c}}$ and maximum spin-up rate between two consecutive time-steps, $\max\{w_{\rm c}\}$, are computed to be $(\sigma_{w_{\rm c}}, \max\{w_{\rm c}\}) = (1.9, 5.7), \, (3.3, 13), \, (3.3, 13)$ for scenarios (1)--(3), respectively. Proton feedback enables the vortices to respond more quickly to spin down, as the pinned flux-tube segments move together with the vortices instead of holding them back. As a result, the differential rotation and the associated Magnus force need not exceed an unpinning threshold, before the neutrons transfer angular momentum to the crust. Furthermore, as shown in Figure \ref{fig:1f} in Section \ref{sec:3.2}, flux tubes bend towards vortices, helping them to latch onto the next flux-tube as they slither. This implies that the distance travelled by vortex segments, and hence the angular momentum transferred to the crust, are smaller with proton feedback.

\subsection{Pinning strength and retarded spin down}
\label{sec:4.2}
\begin{figure}
    \begin{subfigure}[b]{\columnwidth}
    	\includegraphics[width=\textwidth]{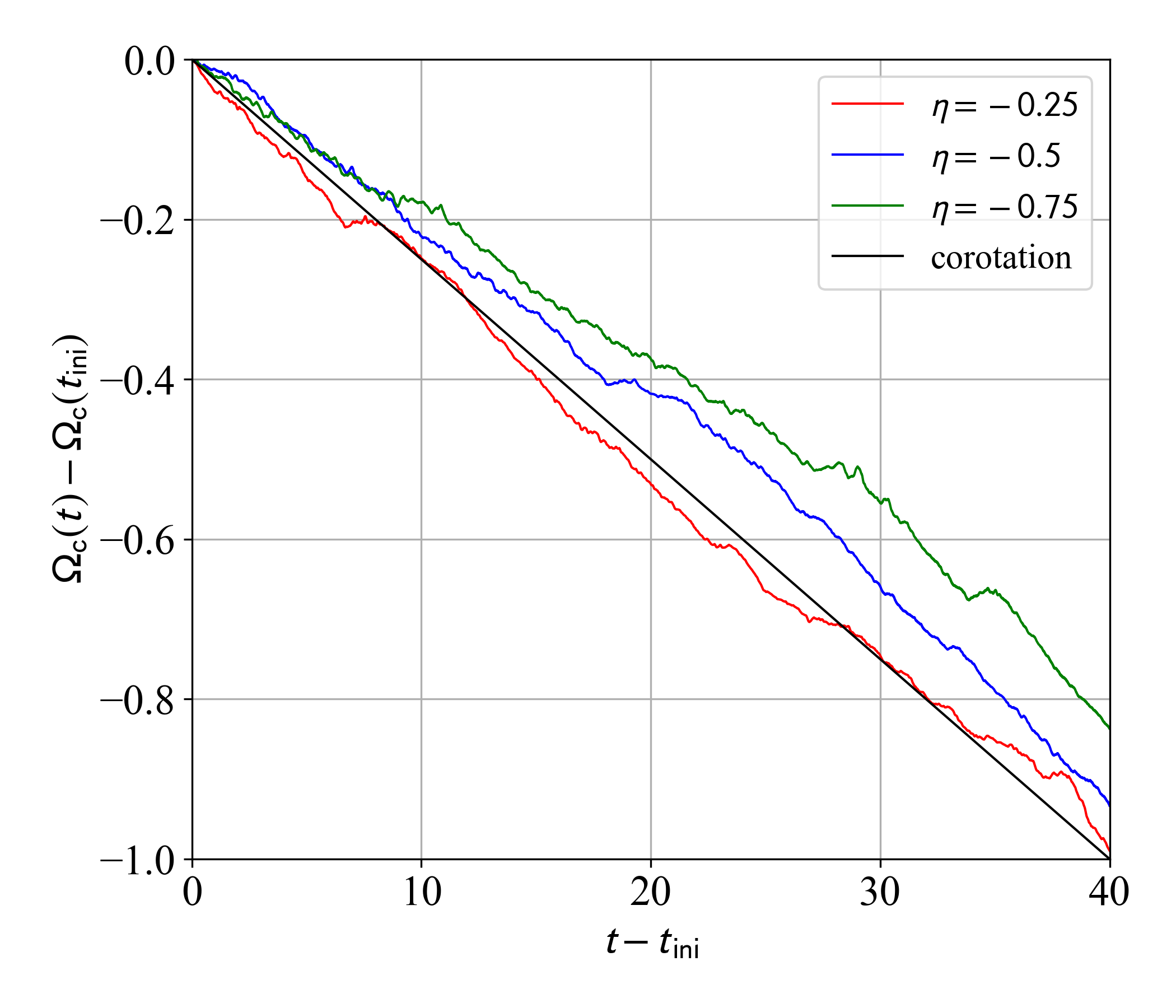}
    	\caption{}
    	\label{fig:5a}
    \end{subfigure}
    \hfill
    \begin{subfigure}[b]{\columnwidth}
    	\includegraphics[width=\textwidth]{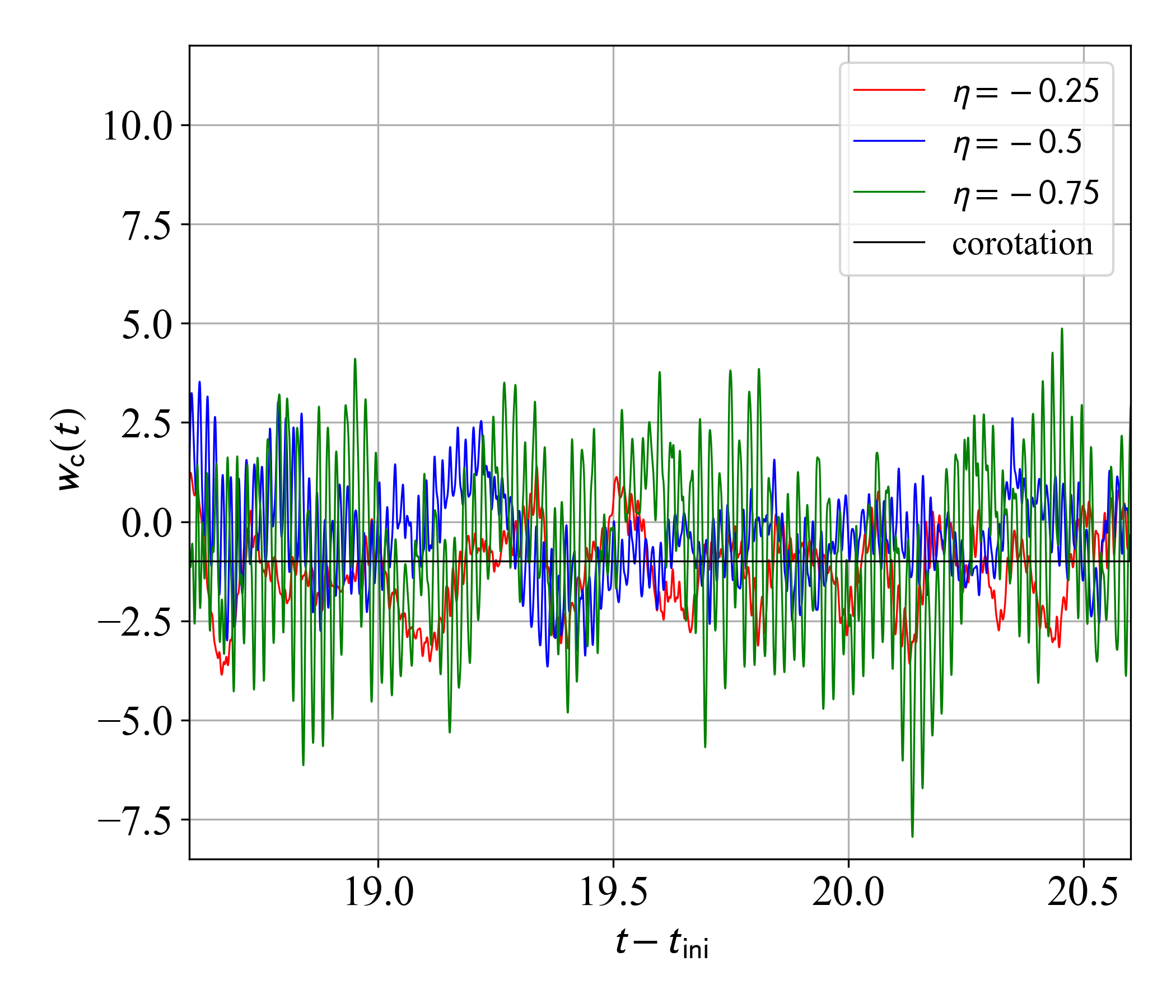}
    	\caption{}
    	\label{fig:5b}
    \end{subfigure}
    \hfill
    \begin{subfigure}[b]{\columnwidth}
    	\includegraphics[width=\textwidth]{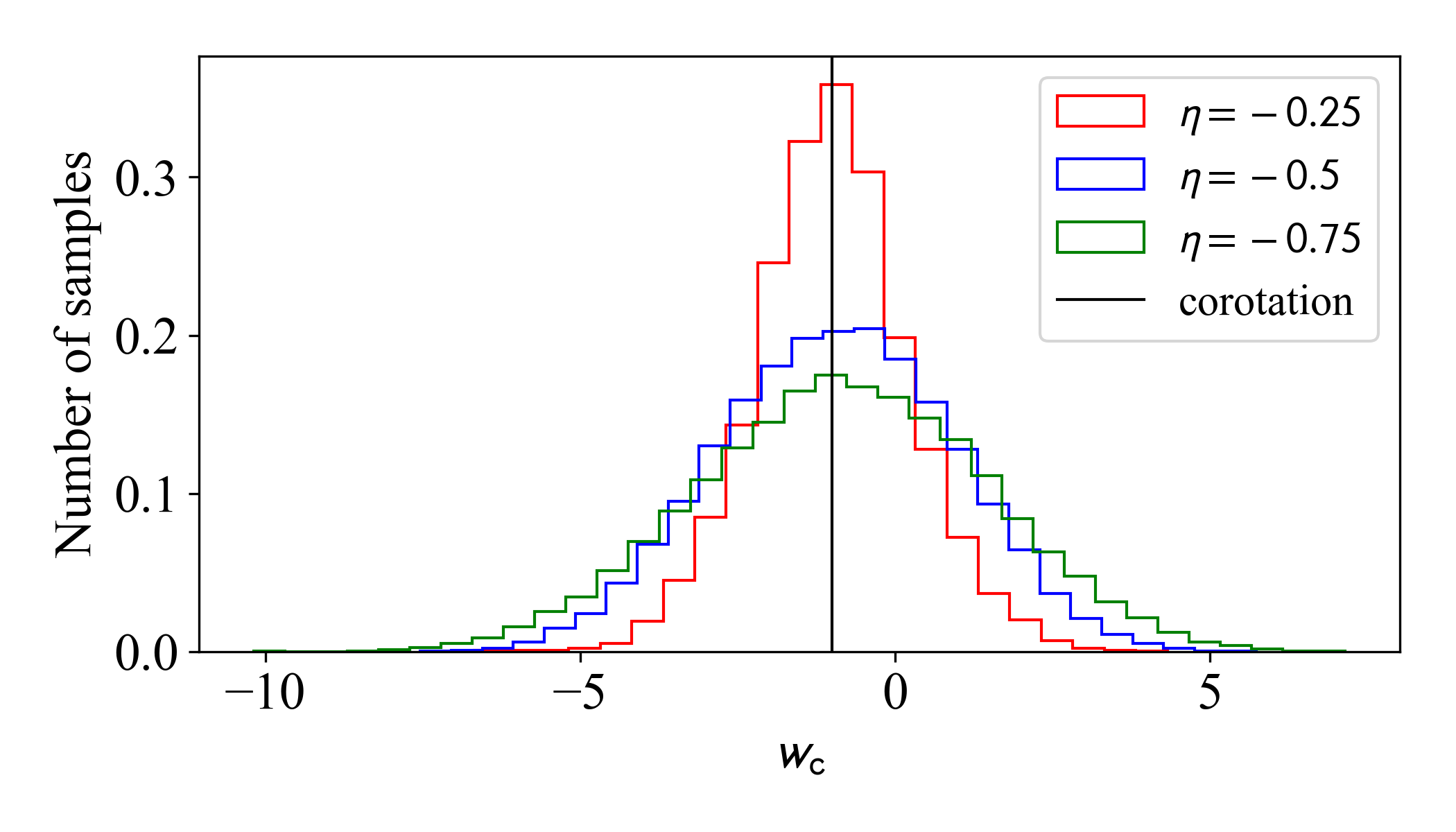}
    	\caption{}
    	\label{fig:5c}
    \end{subfigure}
    \caption{As for scenario (1) in Figure \ref{fig:4}, but with $\eta_{np} = \eta_{pn} = \eta = -0.25 \, ({\rm red}), \, -0.5 \, ({\rm blue}), \, -0.75 \, ({\rm green})$.}
    \label{fig:5}
\end{figure}
\citet{drummond_2018} found that increasing the pinning strength, $|\eta|$, retards the spin down of the crust, when the flux tubes are static. What happens when the protons are free to respond, and the flux tubes are mobile? In this section, we examine this question by running and comparing simulations like those in section \ref{sec:4.1} for $\eta$ in the range $0.25 \leq |\eta| \leq 0.75$. In Figure \ref{fig:5}, we plot $\Omega_{\rm c}(t)  - \Omega_{\rm c}(t_{\rm ini})$ and $w_{\rm c}(t)$ for $t_{\rm ini} \leq t \leq t_{\rm end}$, and the probability density function of $w_{\rm c}(t)$, just like in Figure \ref{fig:4} except for three different values of $\eta_{np} = \eta_{pn} = \eta$. We find that with proton feedback and mobile flux tubes, the crust spins down slower as $|\eta|$ increases. Specifically, we find $\overline{w}_{\rm c}(t_{\rm end}) = -0.99,\, -0.93,\, -0.84$ for $\eta = -0.25,\, -0.5,\, -0.75$, respectively.

The amplitude of the oscillations described in Section \ref{sec:4.1} increases, as $|\eta|$ increases. We see this in the oscillations of $w_{\rm c}(t)$ in Figure \ref{fig:5b}. The standard deviation and maximum spin-up rate between two consecutive time-steps are computed to be $(\sigma_{w_{\rm c}}, \max\{w_{\rm c}\}) = (1.2, 4.3), \, (1.9, 5.7), \, (2.3, 7.2)$, for $\eta = -0.25,\, -0.5,\, -0.75$, respectively. Similarly to \citet{drummond_2018}, we find that the variance of $w_{\rm c}$ increases with $|\eta|$. With stronger pinning, once the Magnus force exceeds the pinning force, vortices unpin with larger initial velocities, and the superfluid transfers a larger amount of angular momentum to the crust. It is interesting that  $\max\{w_{\rm c}\} = 7.2$ for $\eta = -0.75$ with proton feedback is still smaller than $\max\{w_{\rm c}\} = 13$ for $\eta = -0.5$ without proton feedback [scenario (3) in Section \ref{sec:4.1}]. Even with a stronger pinning strength, the maximum spin-up rate of the crust is suppressed by proton feedback, as a result of vortex slithering.

We compare the above results to simulations without proton feedback. With static flux tubes, one obtains $\overline{w}_{\rm c}(t_{\rm end}) = -1.0,\, -0.94,\, -0.90$ for $\eta = -0.25,\, -0.5,\, -0.75$, respectively. That is, stronger pinning to static flux tubes retards the spin down of the crust. All else being equal, the crust in systems with mobile flux tubes spins down slower than the crust in systems with static flux tubes. The percentage differences of $\overline{w}_{\rm c}(t_{\rm end})$ between the simulations with static and mobile flux tubes equal $-4.0\%, \, -4.5\%,\, -7.3\%$ respectively for the three values of $\eta$ quoted earlier in the paragraph. This agrees with the results in section \ref{sec:4.1}, where proton feedback lets the vortices slither and move more freely than zig-zagging between static flux tubes. The percentage difference increases with stronger pinning, as stronger pinning enables more proton feedback.

In this paper, we are unable to reach $|\eta| \geq 1$, unlike \citet{drummond_2018} who covered the range $-100 \leq \eta \leq -1$. The reason is the following. Except near a filament, the neutron and proton density profiles are approximately quadratic ($\propto V_{\omega}$), and the attractive density coupling acts to limit the radii of both condensates. For numerically tractable values of $V_{\omega}$, the condensate radii are too small to contain even one vortex for $|\eta| > 1$. The problem is absent in \citet{drummond_2018}, where only the neutrons are trapped harmonically; the radius of the neutron condensate does not decrease, as $|\eta|$ increases. To circumvent the problem, one can adjust $V_{\omega}$ and $f$, but doing so leads unavoidably to unfavorable adjustments in other properties of the system (see Appendix \ref{appendix:a}, for example), once computational limitations are taken into account. We leave such explorations to future work, when additional computational resources are available. 

\subsection{Misalignment of the rotation and the magnetic axes}
\label{sec:4.3}
In this section, we analyze how the spin down of the crust is affected by the inclination angle, $\theta$, with and without proton feedback. \citet{drummond_2018} found that with static flux tubes, the crust spins down slower for $\theta \neq 0$ than for $\theta = 0$, as vortices slip past the flux tubes more easily. This is not the case with proton feedback. Flux tubes are now mobile and move together with vortices, as discussed in Section \ref{sec:3.1}. Consequently, does tangled vorticity still facilitate the outward motion of vortices and the deceleration of the crust for $\theta \neq 0$, or is it an irrelevant factor? We examine this question in this section by running and comparing simulations like those in Sections \ref{sec:4.1} and \ref{sec:4.2}, but with $0 \leq \theta \leq \pi/2$. 

\begin{figure}
	\includegraphics[width=\columnwidth]{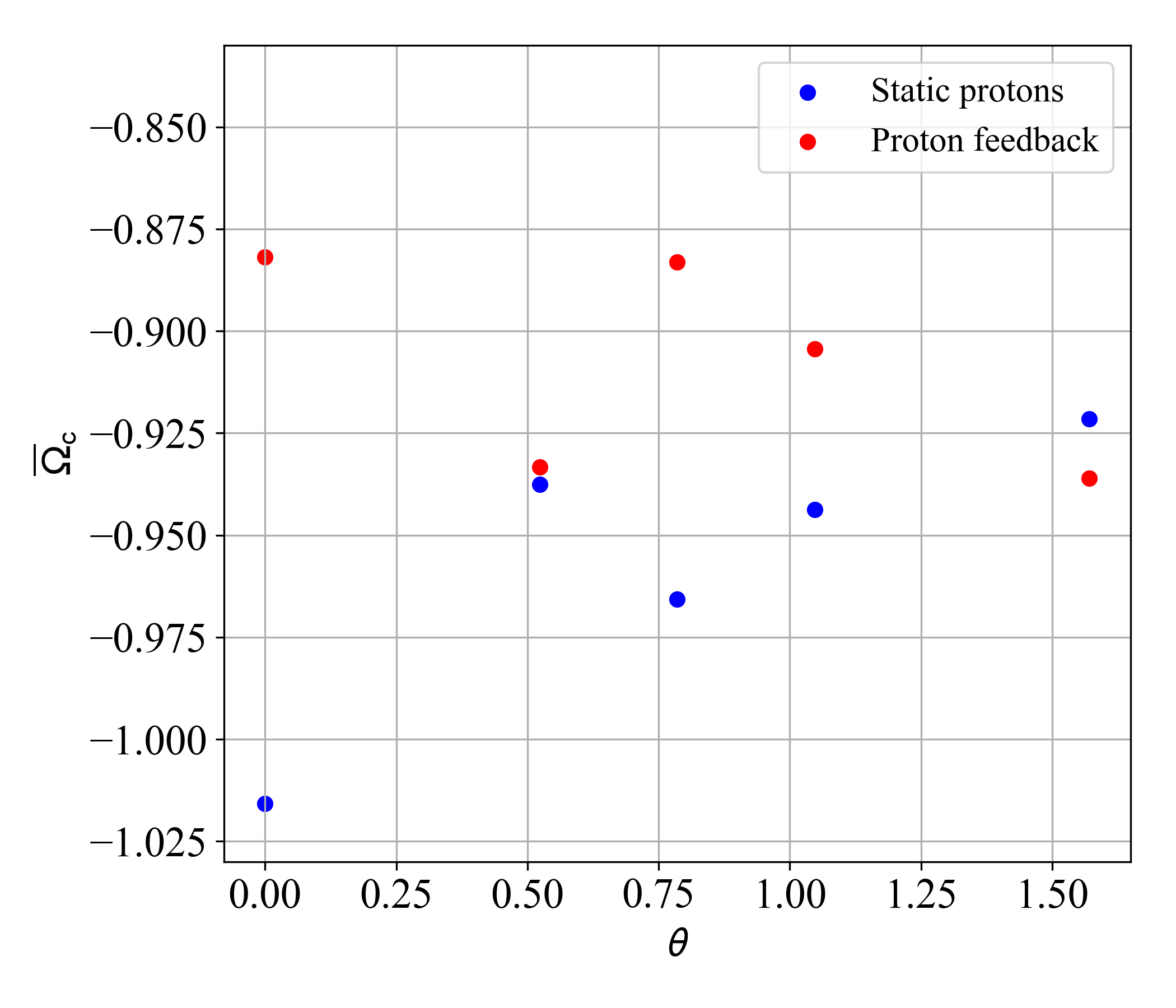}
	\caption{Time-averaged dimensionless spin-down rate, $\overline{w}_{\rm c}(t_{\rm end})$, versus $\theta$, for simulations with static protons (blue symbols) and proton feedback enabled (red symbols). Parameters are the same as in scenario (1) in Figure \ref{fig:4}, but with $0 \leq \theta \leq \pi/2$.}
	\label{fig:6}
\end{figure}

Figure \ref{fig:6} plots $\overline{w}_{\rm c}(t_{\rm end})$ as a function of $0\leq \theta\leq \pi/2$ for simulations with static protons (blue dots) and proton feedback (red dots). Overall, $\overline{w}_{\rm c}(t_{\rm end})$ is higher with proton feedback than without, but interestingly there is no monotonic trend with $\theta$. At $\theta=0$, proton feedback reduces the spin-down rate unambiguously by $\approx 13 \%$, for the reasons discussed in Section \ref{sec:4.1}. By contrast, for $\theta > 0$, the reduction is smaller; $\overline{w}_{\rm c}(t_{\rm end})$ differs by less than $8.5\%$ with and without proton feedback. It is therefore fair to conclude provisionally that tangled vorticity matters less for retarding the spin down of the crust, when proton feedback is included, although larger simulations with higher $N_{\rm v}$ are needed to settle the question confidently. In a neutron star, where one has $N_{\rm f}/N_{\rm v} \sim 10^{14}$, as compared to $N_{\rm f}/N_{\rm v} \sim 10$ in this paper, vortex-flux-tube intersections are more prevalent than in our simulations.

The simulations above are idealized models of a small volume of the outer core of a neutron star. Since the magnetic field in the interior of neutron stars depends on its formation history and is likely to be complicated, it may not make sense to treat the mean field $\overline{\mathbf{B}}$ as uniform, and hence $\theta$ may not be defined uniquely in practice, even though any static magnetic field has a unique magnetic dipole axis in principle. That is, the outer core may divide into zones, and $\theta$ may effectively take different values in different zones. If parts of the outer core with $\theta \approx 0$ decelerate at different rates compared to parts with $\theta \neq 0$, the resulting differential rotation may destabilize the condensates. This interesting possibility is alleviated somewhat by proton feedback, but it may still occur and deserves study in future work.

\section{Tangled filament geometry}
\label{sec:5}
\begin{figure}
    \begin{subfigure}[b]{\columnwidth}
    	\includegraphics[width=\textwidth]{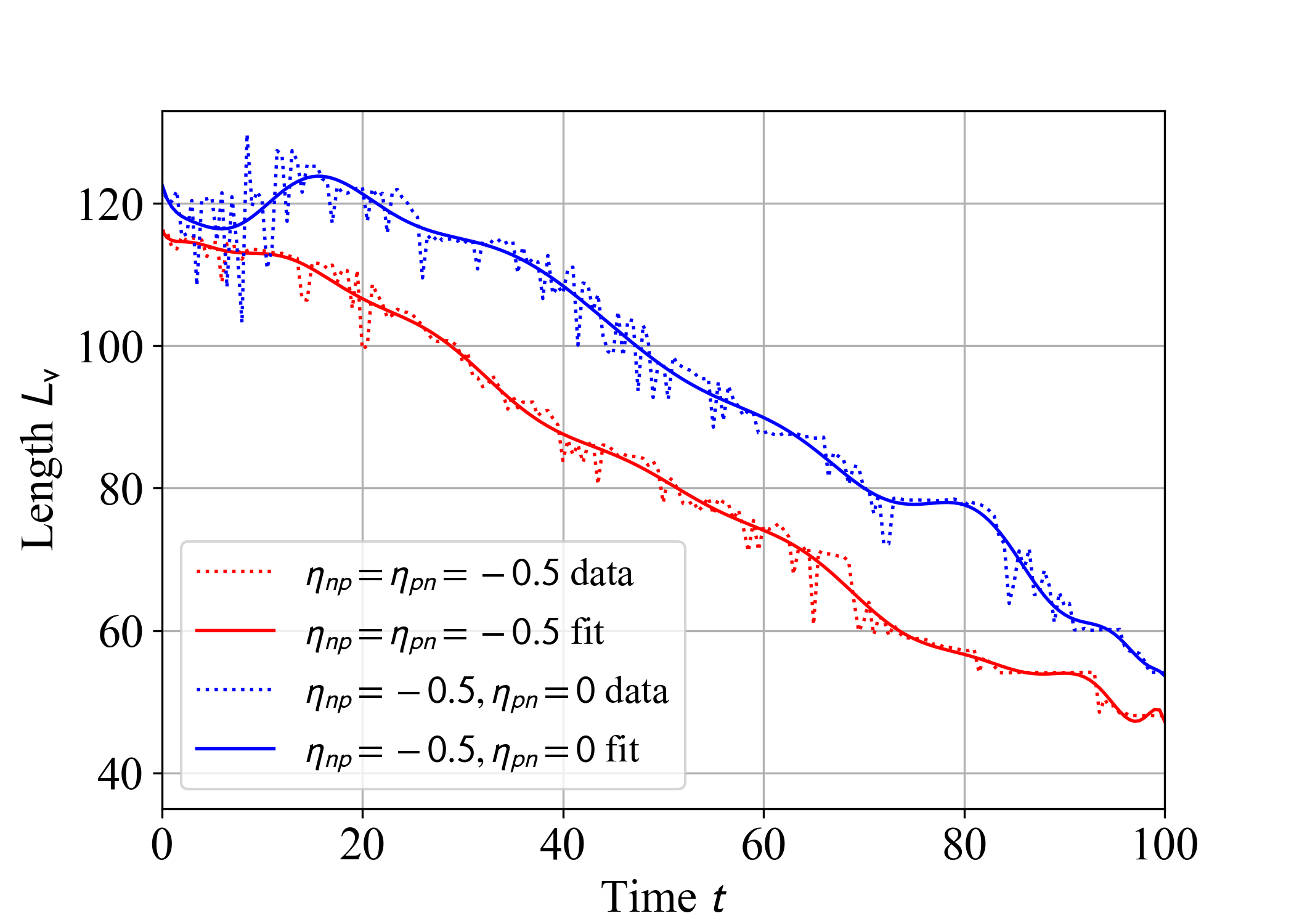}
    	\caption{}
    	\label{fig:7a}
    \end{subfigure}
    \hfill
    \begin{subfigure}[b]{\columnwidth}
    	\includegraphics[width=\textwidth]{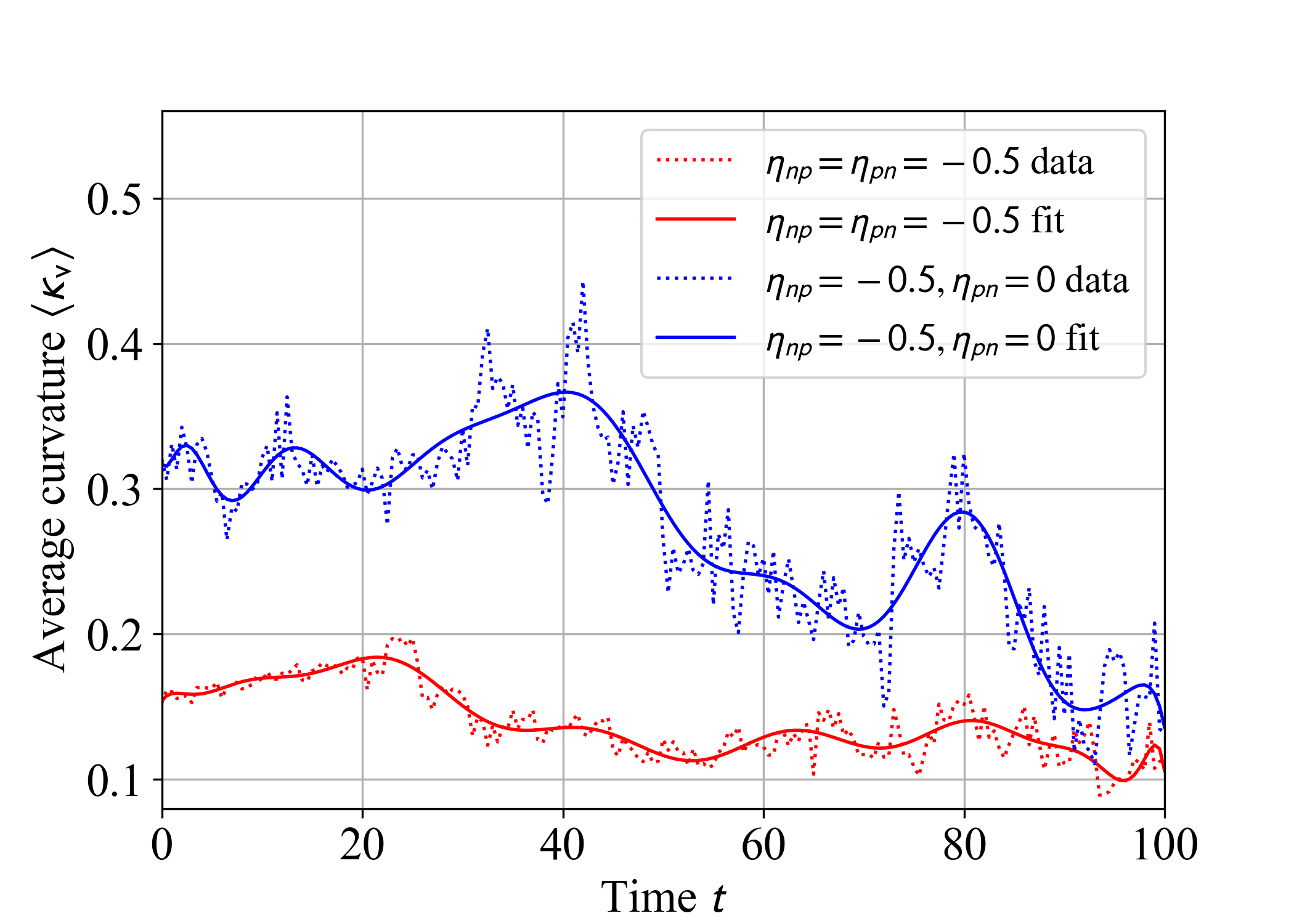}
    	\caption{}
    	\label{fig:7b}
    \end{subfigure}
    \hfill
    \begin{subfigure}[b]{\columnwidth}
    	\includegraphics[width=\textwidth]{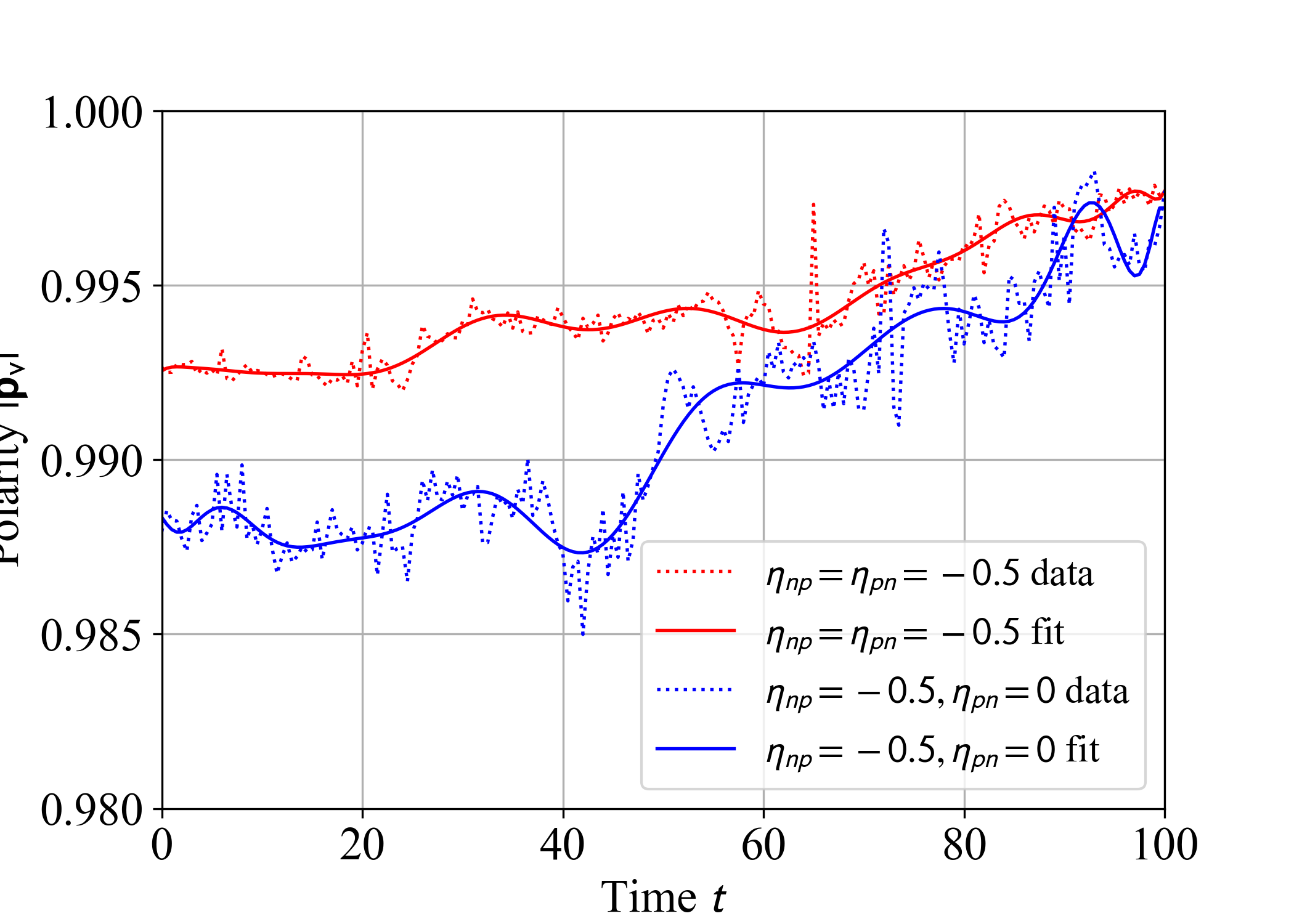}
    	\caption{}
    	\label{fig:7c}
    \end{subfigure}
    \caption{Geometric properties of tangled neutron vortices versus real time $t$ during spin down. (a) Total length $L_{\rm v}$. (b) Length-averaged curvature $\langle \kappa_{\rm v} \rangle$. (c) Magnitude of the length-averaged polarity, $|\mathbf{p}_{\rm v}|$. Results are plotted for systems with (red curve) and without (blue curve) proton feedback (see values of $\eta_{np}$ and $\eta_{pn}$ in legend). Simulation data (dotted curve) and polynomial best fits (solid curves) are plotted. Length units and parameters for the red and blue curves are the same as for scenarios (1) and (2) respectively in Figure \ref{fig:4}.}
    \label{fig:7}
\end{figure}

In this section, we study the effects of proton feedback on the geometrical evolution of the tangled filaments as the crust spins down. Vortices are globally polarised by rotation around $\mathbf{\Omega_{\rm c}}$, which promotes order, but are locally tangled with nearby flux tubes, which generates a superfluid flow parallel to $\mathbf{\Omega_{\rm c}}$ and excites Kelvin waves and hence disorder \citep{drummond_2018}. For spin down without proton feedback, an instability arises as outward-moving vortices push against rigid flux tubes, creating unpolarised grid turbulence \citep{link_instability_2012}. With proton feedback, does one expect less turbulence, as the flux tubes are no longer rigid? We explore this question by measuring three essential geometric quantities of tangled vortices during spin down of the crust, namely the total length 
\begin{equation}
\label{eq:5.15}
    L_{\rm v} = \int{d\xi \ |\mathbf{s}_{\rm v}'(\xi)|},
\end{equation}
the length-averaged curvature 
\begin{equation}
\label{eq:5.16}
    \langle \kappa_{\rm v} \rangle = \frac{1}{L_{\rm v}} \int{d\xi \  \frac{|\mathbf{s}_{\rm v}'(\xi) \times \mathbf{s}_{\rm v}''(\xi)|}{|\mathbf{s}_{\rm v}'(\xi)|^2}},
\end{equation}
and the length-averaged polarity
\begin{equation}
\label{eq:5.17}
    \mathbf{p}_{\rm v} = \frac{1}{L_{\rm v}} \int{d\xi \ \mathbf{s}_{\rm v}'(\xi)},
\end{equation}
where $\mathbf{s}_{\rm v}(\xi)$ is the displacement of a point on the vortex filament from the origin, labelled by an affine parameter $0 \leq \xi \leq 1$. We present the three quantities above (dotted curves) and associated polynomial best fits (solid curves) versus real-time in Figure \ref{fig:7}, for systems with (red curve) and without (blue curve) proton feedback, like scenarios (1) and (3) respectively in Section \ref{sec:4.1}. We identify the vortices in our solutions using a vortex finding algorithm, described in detail in Section 5 of \citet{drummond_2018}.

The geometric properties of tangled flux tubes are also of astrophysical interest. The magnetic field structure on microscopic scales may play a role in determining the magnetic field structure on macroscopic scales (and vice versa), possibly with observable consequences (e.g., slow wandering of the magnetic poles or the dipole-to-quadrupole ratio) \citep{macy_pulsar_1974,ruderman_neutron_1998,barsukov_spin_2013}. We are unable to study this important question fully in this paper because we assume $\overline{\mathbf{B}} = \nabla \times \overline{\mathbf{A}}$ is static. However, we take a first step by characterizing the geometric properties of the tangled flux tubes by calculating the analogs of (\ref{eq:5.15})--(\ref{eq:5.17}) for the protons. The results are summarized briefly in Appendix \ref{appendix:c}.

\subsection{Length}
\label{sec:5.1}
Before the braking torque turns on at $t=0$, Figure \ref{fig:7a} shows that $L_{\rm v}$ of a vortex pinned to rigid flux tubes (blue curve) is $5.2\%$ greater than for a vortex pinned to mobile flux tubes (red curve). With proton feedback, flux tubes are free to tangle with vortices (see Figure \ref{fig:1}), so the vortex and its nearby flux tubes share the task of adjusting to local stresses, and the vortex stretches less than it would otherwise. After the braking torque turns on ($t>0$), vortex length without proton feedback (blue curve in Figure \ref{fig:7a}) alternately decreases at $t \approx 5$ and increases at $t \approx 15$ as Kelvin waves are generated, as also observed by \citet{drummond_2018}. In contrast, $L_{\rm v}$ fluctuates less with time, when proton feedback is introduced (red curve in Figure \ref{fig:7a}), and there are no distinct peaks and troughs visible at $t \approx 5$ and $t \approx 15$ for example. Kelvin waves and other perturbations appear to be quenched by proton feedback. Overall, as $t$ increases, vortices annihilate progressively at the crust, and $L_{\rm v}$ decreases with and without proton feedback.

\subsection{Curvature}
\label{sec:5.2}
Figure \ref{fig:7b} plots the length-averaged curvature $\langle \kappa_{\rm v} \rangle$ versus time. $\langle \kappa_{\rm v} \rangle$ measures how bent the vortices are; the average radius of curvature equals $\langle \kappa_{\rm v} \rangle^{-1/2}$. At equilibrium ($t=0$), vortices are $52\%$ less curved with proton feedback than without, for the reason explained in Section \ref{sec:5.1}. The creation of Kelvin waves causes $\langle \kappa_{\rm v} \rangle$ to increase initially, before $\langle \kappa_{\rm v} \rangle$ slowly decays. The fluctuations in $\langle \kappa_{\rm v} \rangle$ are from vortex-vortex, vortex-flux-tube and vortex-boundary interactions. The fluctuation amplitude is smaller with proton feedback than without; the standard deviation in $\langle \kappa_{\rm v} \rangle$ equals $2.4 \times 10^{-2}$ and $6.9 \times 10^{-2}$ respectively from Figure \ref{fig:7b}. In other words, the neutron superfluid is less turbulent during spin down with proton feedback than without. We caution that microscopic turbulence in the vortex tangle does not translate automatically to macroscopic ``jitter'' in spin down of the crust. Although it is true that the crust spins down more smoothly with proton feedback, as noted in Section \ref{sec:4.1}, it is hard to ascertain how much of this behaviour is caused by the lower value of $\langle \kappa_{\rm v} \rangle$ (as opposed to being correlated with it).

\subsection{Polarity}
\label{sec:5.3}
Figure \ref{fig:7c} plots the magnitude of the length-averaged polarity $|\mathbf{p}_{\rm v}|$ versus time. $|\mathbf{p}_{\rm v}|$ measures how aligned the vortices are, with $|\mathbf{p}_{\rm v}| = 1$ indicating perfect global alignment and $|\mathbf{p}_{\rm v}| = 0$ indicating random alignment (isotropic tangle). At equilibrium ($t = 0$), we find that vortices are $0.47\%$ more polarised with proton feedback than without. This accords with the vortices being marginally less stretched (Section \ref{sec:5.1}) and less curved (Section \ref{sec:5.2}). After the braking torque is applied ($t>0$), the fluctuations in $|\mathbf{p}_{\rm v}|$ (like those in $L_{\rm v}$ and $\langle \kappa_{\rm v} \rangle$) are of smaller amplitude with proton feedback than without; the standard deviation in $|\mathbf{p}_{\rm v}|$ equals $1.7 \times 10^{-3}$ and $3.3 \times 10^{-3}$ respectively from Figure \ref{fig:7c}, consistent with vortex slithering supplanting vortex zig-zagging (see Section \ref{sec:3.2}). As $t$ increases, $\langle \kappa_{\rm v} \rangle$ slowly decreases while $|\mathbf{p}_{\rm v}|$ slowly increases in both scenarios; $N_{\rm v}$ halves approximately from $t = 0$ to $t=100$, the average vortex separation increases, vortex-vortex repulsion decreases, and vortices align more easily with $\mathbf{\Omega}_{\rm c}$. The secular change in $\langle \kappa_{\rm v} \rangle$ and $|\mathbf{p}_{\rm v}|$ is associated with vortices exiting the system. However, $N_{\rm v}$ remains constant and large between real glitches in a neutron star, so we do not expect the secular trend in Figure \ref{fig:7c} to have astrophysical consequences.

\section{Conclusions}
In this paper, we conduct three-dimensional simulations of dynamic, tangled neutron vortices and proton flux tubes, as an idealized microscopic description of the magnetized, superfluid, and superconducting outer core of a neutron star. The simulations solve the GPE and TDGLE simultaneously and therefore include feedback from the protons on the neutrons and vice versa, extending previous calculations where such feedback is missing, and the flux tubes are static and rectilinear \citep{drummond_2017,drummond_2018}. We first solve the GPE and TDGLE in imaginary time to find the equilibrium state, as a function of $\eta$ and $\theta$. Then, we apply a braking torque to the crust and evolve the system in real time to study the effects of proton feedback on the spin down of the crust and on the stability of the vortex tangles.

In the ground state, for $\theta \neq 0$, proton feedback decreases the length and curvature and increases the polarity of the vortex filaments, because flux tubes bend towards vortices in response to local Magnus and pinning forces (see Section \ref{sec:5}). However, the neutron vortices remain tangled, in response to competition between local Magnus and pinning forces and the global orientations of the rotation and magnetic axes. During spin down, a new type of vortex-flux-tube motion is observed. For $\theta = 0$, vortices pair up rectilinearly and travel cylindrically outwards while pinned to flux tubes along their entire length (see Section \ref{sec:3.1}). For $\theta \neq 0$, vortex-flux-tube intersections occur along segments rather than the entire length, and vortices partially pin to different flux tubes (see Section \ref{sec:3.2}). Similarly to $\theta = 0$, vortex segments for $\theta \neq 0$ move together with pinned flux-tube segments without unpinning. This causes smoother and retarded spin down of the crust and smaller rotational glitches (see Section \ref{sec:4.1}). The retardation of the spin down of the crust increases with increasing $|\eta|$, as proton feedback becomes more significant (see Section \ref{sec:4.2}). In section \ref{sec:4.3}, we find that, compared to the pinning of vortices to static flux tubes, the pinning to mobile flux tubes causes the time-averaged spin-down rate of the crust to depend less on tangled vorticity ($\theta \neq 0$). In section \ref{sec:5},  we find that proton feedback quenches the oscillations in $L_{\rm v}$, $\langle \kappa_{\rm v} \rangle$ and $|\mathbf{p}_{\rm v}|$ of the vortex tangles, as flux tubes follow and damp the oscillations.

We remind the reader that the model in this paper is highly idealised for the following reasons. (i) Due to numerical limitations, the parameters used here are not in the same regime as those in neutron stars, although every effort is made to preserve the ordering (albeit not the dynamic range) of dimensionless parameters, e.g. in a neutron star, we have $N_{\rm v} \sim 10^{16} \ll N_{\rm f} \sim 10^{30}$, which compares to $N_{\rm v} \sim 22 \ll N_{\rm f} \sim 200$ in the simulations. (ii) Larger $N_{\rm v}$ and $N_{\rm f}$ are needed to study whether large-scale avalanches triggered by vortex unpinning are suppressed or promoted by proton feedback. (iii) The time-dependent GPE and TDGLE are only approximate descriptions for a high-density, strongly interacting neutron superfluid and a proton superconductor in a neutron star. (iv) Setting $\mathbf{B} = \overline{\mathbf{B}}$ is another approximation. 

Many interesting questions remain. For example, how do the dynamics change, if the magnetic field evolves self-consistently? Is the flux-tube tangle susceptible to magnetized versions of the Donnelly-Glaberson instability \citep{link_instability_2012,van_eysden_hydrodynamic_2018}? Are there observational consequences, if new, interfacial dynamics emerge at the boundaries between different superfluid and superconducting phases confined to ``domains'' within the star, e.g. in a type-1.5 superconductor  \citep{haber_critical_2017,Wood_2022}, as happens in experiments with liquid helium \citep{blaauwgeers_shear_2002}? These and other puzzles may be investigated profitably within the GPE-TDGLE framework described in this paper.

\section*{Acknowledgements}

The authors acknowledge helpful discussions with Thippayawis Cheunchitra. Parts of this research are supported by the University of Melbourne Science Graduate Scholarship --- 2021, Australia and the Australian Research Council (ARC) Centre of Excellence for Gravitational Wave Discovery (OzGrav) (grant number CE170100004). This research was supported by the University of Melbourne's Research Computing Services and the Petascale Campus Initiative.

\section*{Data Availability}

Simulation data and code used in this paper can be made available upon request by emailing the corresponding author.



\bibliographystyle{mnras}
\bibliography{mnras_template} 




\appendix

\section{Condensate diameter}
\label{appendix:a}

\begin{figure}
    \centering
    \begin{subfigure}[b]{0.86\columnwidth}
    	\includegraphics[width=\textwidth]{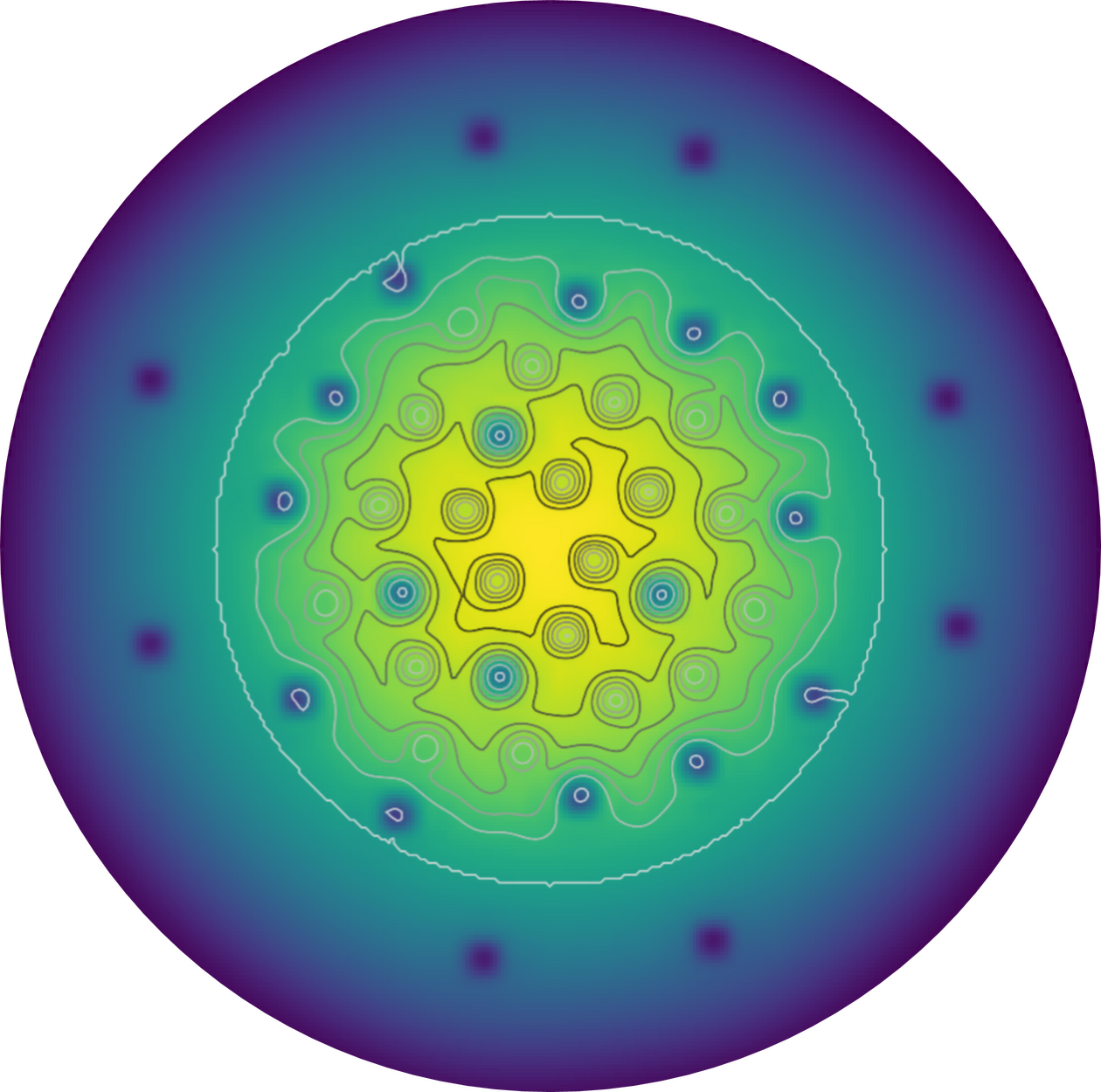}
    	\caption{}
    	\label{fig:8a}
    \end{subfigure}
    \par\bigskip
    \begin{subfigure}[b]{0.86\columnwidth}
    	\includegraphics[width=\textwidth]{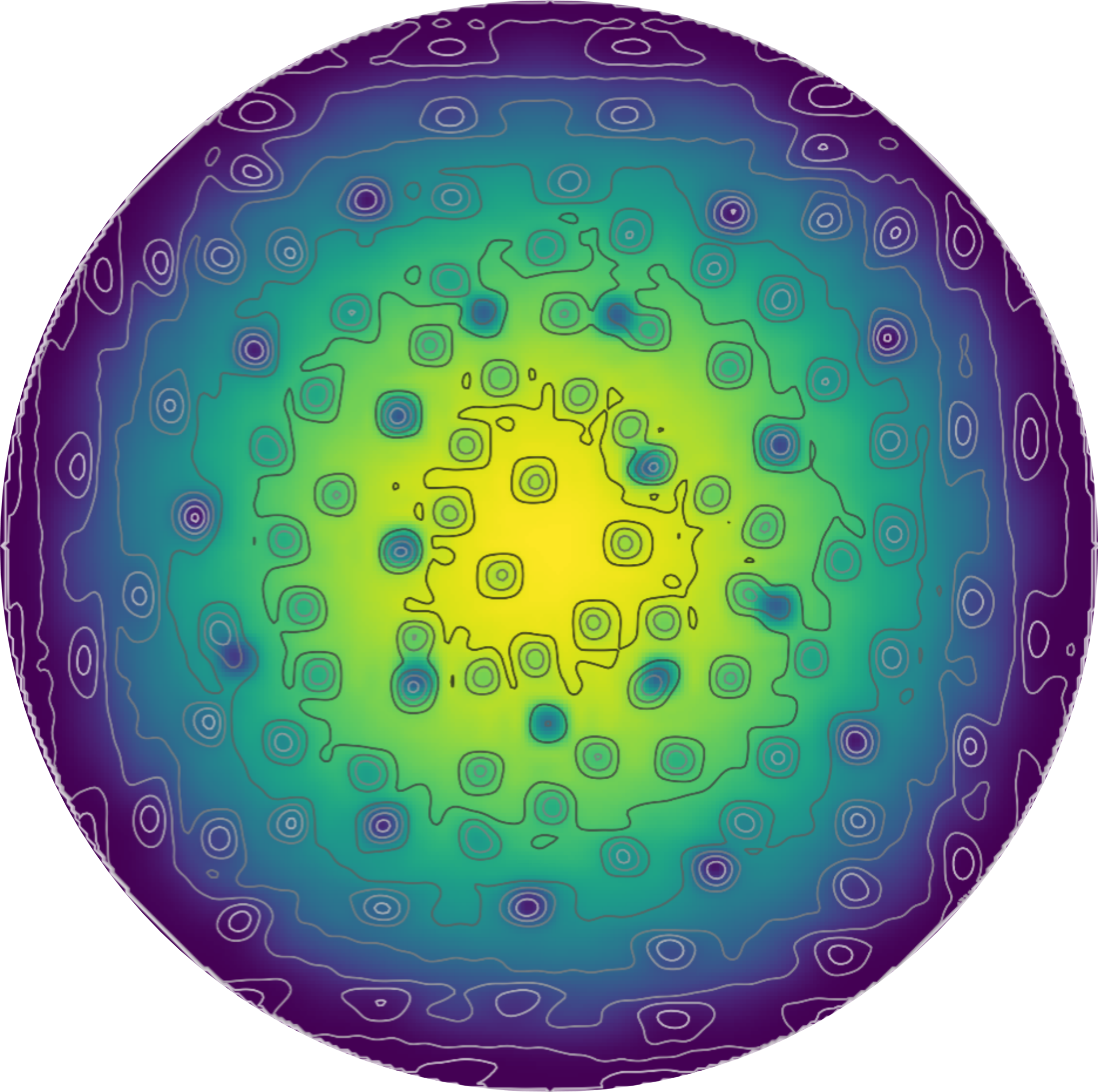}
    	\caption{}
    	\label{fig:8b}
    \end{subfigure}
    \hfill
    \caption{Condensate diameters versus $f$; snapshots of a two-dimensional slice of the ground states of $|\psi|^2$ (coloured plot) and $|\phi|^2$ (contour plot). The parameters are the same as the $\theta = 0$ simulation in Figure \ref{fig:6}, except for (a) $f=0.05$, and (b) $f=1.25$.}
    \label{fig:8}
\end{figure}
In the outer core of a neutron star, the proton number density is $\lesssim 5\%$ of the neutron number density \citep{chau_implications_1992}. In this paper, setting the proton fraction to be small, i.e.\ $f \ll 1$, results in the proton condensate having a smaller radius than the neutron condensate. This is because the neutrons and protons reside in the same harmonic trap (gravitational potential well of the star), so fewer particles exert less pressure. As a result, the regime $f \ll 1$ produces regions where there are neutron vortices but no proton flux tubes. For example, for $f = 0.05$ and $\theta = 0$, we plot a two-dimensional slice of the ground state of the neutrons (coloured plot) and protons (contour plot) in Figure \ref{fig:8a}. With this arrangement, the vortices in the outer ring lie outside all the flux tubes and therefore cannot pin to them, defeating the purpose of the simulations in this paper. In comparison, for $f = 1.25$ in Figure \ref{fig:8b}, vortices overlap with flux tubes throughout the condensate, as desired astrophysically.

\section{Evolving the magnetic field}
\label{appendix:b}
The evolution of $\mathbf{A}$ in equation (\ref{eq:2.6}) is governed by Maxwell's equations in dimensionless form, viz. 
\begin{equation}
\label{eq:B1}
    \sigma \frac{\partial \mathbf{A}}{\partial t} = - \nabla \times (\nabla \times \mathbf{A}) - \frac{1}{2\kappa_{\rm GL}}(\phi^*\nabla \phi - \phi \nabla \phi^*) - |\phi|^2 \mathbf{A},
\end{equation}
where $\sigma$ is the dimensionless conductivity of the normal matter (normal protons and relativistic electrons) in the fluid mixture \citep{gor1996generalization,chapman_vortex_1997}. We do not solve equation (\ref{eq:B1}) for the reasons discussed in Section \ref{2.2} and \ref{2.4}. Additionally, the numerical techniques used in this paper \citep{caradoc-davies_vortex_2000} rely on calculating spatial derivatives via Fourier transforms, and therefore assume periodic boundary conditions. Periodic boundary conditions are incompatible with a type-II superconductor; instead, one applies a quasi-periodic boundary condition \citep{wood_quasiperiodic_2019}. We postpone the implementation of these subtleties to future work, when we come to solve (\ref{eq:B1}), noting that fast light-crossing time-scale dynamics in Maxwell's equations can be neglected when studying the slow, spin-down-driven dynamics in this paper.

\section{Statistical properties of tangled flux tubes}
\label{appendix:c}
\begin{figure}
    \begin{subfigure}[b]{\columnwidth}
    	\includegraphics[width=\textwidth]{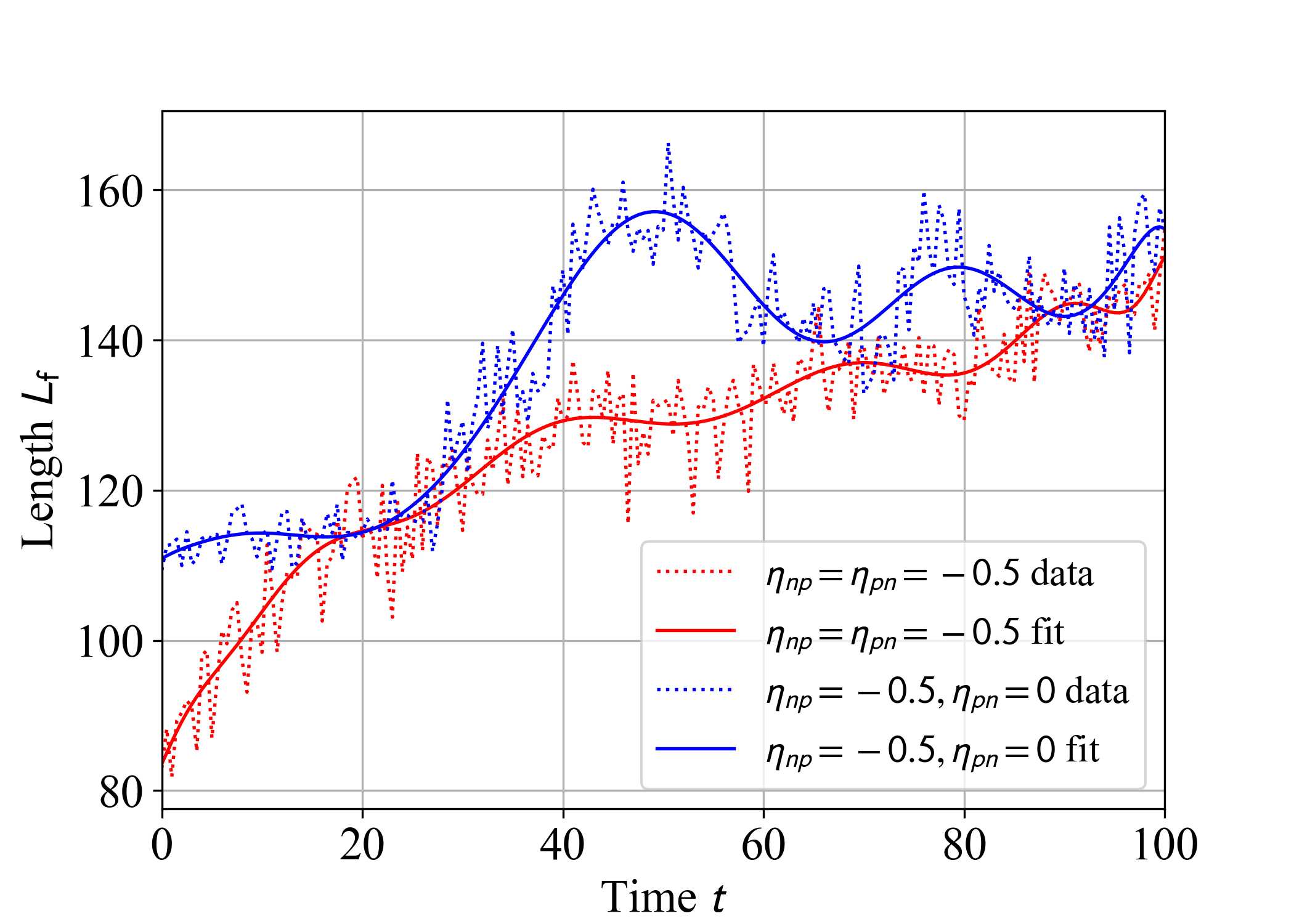}
    	\caption{}
    	\label{fig:9a}
    \end{subfigure}
    \hfill
    \begin{subfigure}[b]{\columnwidth}
    	\includegraphics[width=\textwidth]{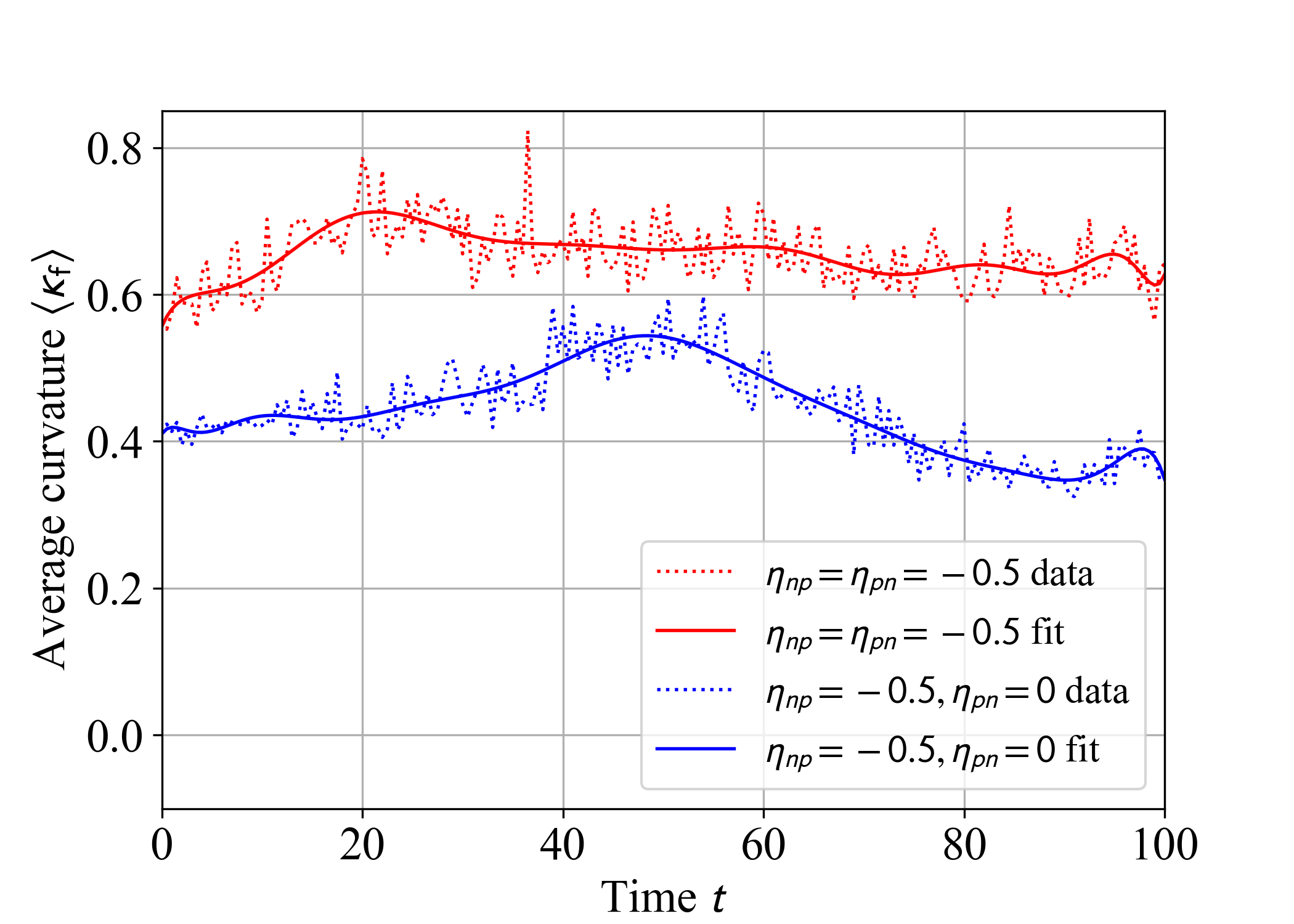}
    	\caption{}
    	\label{fig:9b}
    \end{subfigure}
    \hfill
    \begin{subfigure}[b]{\columnwidth}
    	\includegraphics[width=\textwidth]{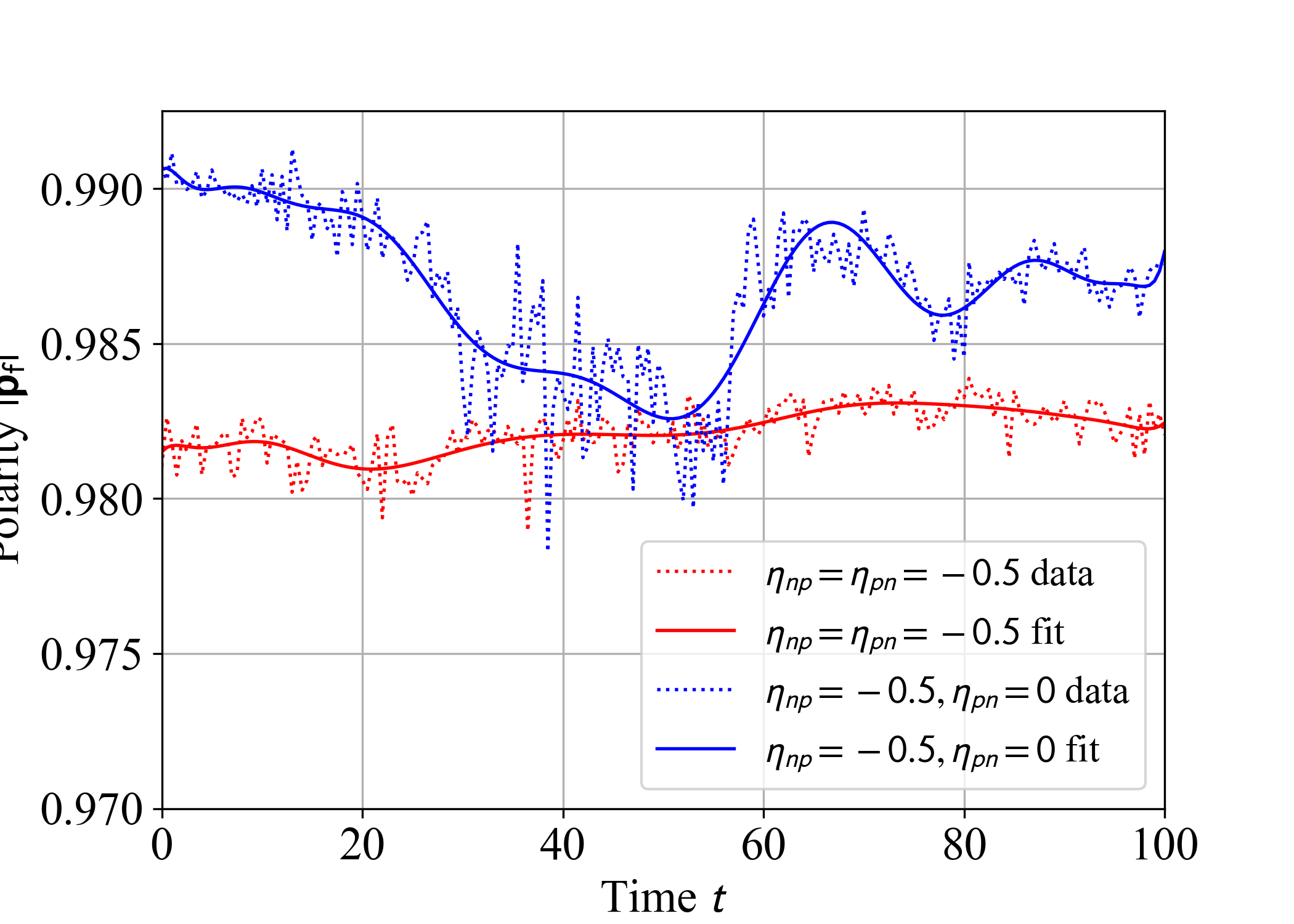}
    	\caption{}
    	\label{fig:9c}
    \end{subfigure}
    \caption{As for Figure \ref{fig:7}, but for flux tubes instead of vortices and with the integrals in (\ref{eq:5.15})--(\ref{eq:5.17}) calculated in a cube with side length $2 \lambda_p$ centred at the origin.}
    \label{fig:9}
\end{figure}
In this appendix, we study the spin-down-driven evolution of the length $L_{\rm f}$, length-averaged curvature $\langle \kappa_{\rm f} \rangle$, and magnitude of the length-averaged polarity $|\mathbf{p}_{\rm f}|$ of the tangled flux tubes. The variables $L_{\rm f}$, $\langle \kappa_{\rm f} \rangle$, and $|\mathbf{p}_{\rm f}|$ are calculated using equations (\ref{eq:5.15})--(\ref{eq:5.17}), with the subscript `v' replaced by the subscript `f' everywhere. The line integrals in equations (\ref{eq:5.15})--(\ref{eq:5.17}) are restricted to a cube with side length $2 \lambda_p$ centered at the origin instead of the whole simulation volume to reduce the computational expense of integrating over all the flux tubes. This restriction is not applied when calculating $L_{\rm v}$, $\langle \kappa_{\rm v} \rangle$, and $|\mathbf{p}_{\rm v}|$, because we have $N_{\rm v} \ll N_{\rm f}$. In fully developed turbulence, $\langle \kappa_{\rm f} \rangle$ and $\mathbf{p}_{\rm f}$ are independent of the box size, because the line integrals are normalized by $L_{\rm f}$ \citep{donnelly_quantized_1991}. 

In Figure \ref{fig:9}, we plot the raw simulation output and polynomial best fits for $L_{\rm f}$, $\langle \kappa_{\rm f} \rangle$, and  $|\mathbf{p}_{\rm f}|$ as functions of time with (red curves) and without (blue curves) proton feedback, i.e.\ the same two scenarios as in Section \ref{sec:5}. In both scenarios, vortices move outwards in response to the deceleration of the crust, and flux tubes are attracted to the rigid array of artificial pinning cylinders emulating flux freezing, as defined by $V_{\rm f}$ in equation (\ref{eq:2.13}); see Section \ref{2.4}. With proton feedback enabled, flux tubes also pin to vortices. The pinning cylinders are parallel to $\overline{\mathbf{B}}$, but the flux tubes globally align with an axis that is inclined between $\overline{\mathbf{B}}$ and $\mathbf{\Omega}_{\rm c}$, as explained in footnote \ref{footnote:2}. As a result, flux tubes tangle around the rigid pinning cylinders whether they pin to vortices or not. As the crust spins down, the angular momentum of the protons decreases, flux tubes leave the condensate, and the remaining flux tubes align more closely with $\overline{\mathbf{B}}$.

Before $N_{\rm em}$ is turned on, the lengths of the flux tubes are $L_{\rm f} = 83$ and $110$ with and without proton feedback respectively (see Figure \ref{fig:9a}). Flux tubes that pin to vortices are shorter than those that pin to the pinning cylinders in (\ref{eq:2.13}). This is because vortices restricted to a cube are shorter than pinning cylinders for $\theta = \pi/6$, and flux-tube-vortex pinning is stronger than the flux-tube-lattice-site pinning, i.e.\ one has $\mathcal{H}_{\rm int}[\phi, \psi] \gg V_{\rm f}$ away from flux tubes and pinning cylinders. The curvature and the magnitude of the polarity of the flux tubes in the ground state, $(\langle \kappa_{\rm f} \rangle,|\mathbf{p}_{\rm f}|)$, equals $(0.56, 0.98)$ and $(0.41, 0.99)$ with and without proton feedback respectively (see Figure \ref{fig:9b} and \ref{fig:9c}). Flux tubes are more tangled and less polarised when pinned to vortices. 

After $N_{\rm em}$ is turned on at $t = 0$, $L_{\rm f}$ slowly increases in both scenarios, as $\Omega_{\rm c}(t)$ decreases, as flux tubes tend to align with $\overline{\mathbf{B}}$. Similarly to vortices, the oscillations in the three geometric quantities in Figure \ref{fig:9} are from flux-tube-flux-tube, vortex-flux-tube and flux-tube-boundary interactions, as well as the flux-tube-pinning-cylinder interactions for the protons. For $\eta_{pn} = 0$ (blue curves), Kelvin waves cause $L_{\rm f}$ and $\langle \kappa_{\rm f} \rangle$ to increase and $|\mathbf{p}_{\rm f}|$ to decrease until they reach their extremas at $t \approx 50$. In comparison, for $\eta_{pn} \neq 0$, the flux-tube oscillations are damped by flux tubes pinning to vortices (red curves). The standard deviations over a simulation run, $(\sigma_{\langle \kappa_{\rm f} \rangle}, \, \sigma_{|\mathbf{p}_{\rm f}|})$, equal $(0.044, 8.6 \times 10^{-4})$ and $(0.063, 2.6 \times 10^{-3})$ with and without vortex pinning respectively. 


\bsp	
\label{lastpage}
\end{document}